\documentclass[journal]{new-aiaa}

\title{Adjoint-based optimization for thrust performance of a three-dimensional pitching-rolling plate}

\author{Min Xu\footnote{Ph.D., Department of Mechanical and Aerospace Engineering}}
\affil{New Mexico State University, Las Cruces, NM 88001}
\author{Mingjun Wei\footnote{Associate Professor, Department of Mechanical and Nuclear Engineering, AIAA Associate Fellow}}
\affil{Kansas State University, Manhattan, KS 66506}
\author{Chengyu Li\footnote{Assistant Professor, Department of Mechanical Engineering, AIAA member}}
\affil{Villanova University, Villanova, PA 19085}
\author{Haibo Dong\footnote{Associate Professor, Department of Mechanical and Aerospace Engineering, AIAA Associate Fellow}}
\affil{University of Virginia, Charlottesville, Virginia 22904}

\usepackage{amsmath,amssymb,epsfig,pifont,color,colordvi,float,verbatim}
\usepackage{hyperref,graphicx}
\usepackage{longtable,tabularx}

\def\includefigs{\let\ifincfigs=\iftrue}
\def\noincludefigs{\let\ifincfigs=\iffalse}
\includefigs

\newbox\epsfvertlab
\newbox\epsfhorlab
\newbox\epsffiglab

\newdimen\epsfvlabsize
\newdimen\scott

\def\setvlabel#1{\setbox\epsfvertlab=\vbox{\hbox{#1}}}%
\def\sethlabel#1{\setbox\epsfhorlab=\vbox{\hbox{#1}}}%
\def\figlab#1 #2 #3{\setbox\epsffiglab=\vbox to 0pt{%
\ifvoid\epsffiglab\else\box\epsffiglab\fi\vss\hbox to 0pt{\raise #2 \hbox{\hskip #1 #3}\hss}}}

\newdimen\fighor
\newdimen\figver
\newbox\rotbox
\long\def\lrlap#1{\hbox to 0pt{#1\hss}}
\long\def\verttex#1#2#3{{\fighor = #1\figver = #2\vbox to \figver{\vss%
\hbox to \fighor{\hfill\hsize=\fighor%
\lrlap{\rotstart{-90 rotate}\vbox to \fighor{#3\vfil}\rotfinish}}}}}

\def\dvipsvspec#1{\special{ps:#1}}
\def\dvipsrotstart#1{\dvipsvspec{gsave currentpoint currentpoint translate
   #1 neg exch neg exch translate}}
\def\dvipsrotfinish{\dvipsvspec{currentpoint grestore moveto}}

\def\rotstart#1{\dvipsrotstart{#1}}
\def\rotfinish{\dvipsrotfinish}

\def\epsfsetlab{%
\ifvoid\epsfvertlab%
\else%
\verttex{\epsfvlabsize}{\epsfysize}%
{\hbox to \epsfysize{\hss\box\epsfvertlab\hss}}%
\fi%
\ifvoid\epsfhorlab%
\else%
\scott=\epsfxsize%
\advance\scott by \epsfvlabsize%
\rlap{\vtop{\hrule height0pt\hbox to \scott{\hss\box\epsfhorlab\hss}}}%
\fi%
}

\def\epsfsetover{\ifvoid\epsffiglab\else\box\epsffiglab\fi}

\newread\epsffilein    
\newif\ifepsffileok    
\newif\ifepsfbbfound   
\newif\ifepsfverbose   
\newdimen\epsfxsize    
\newdimen\epsfysize    
\newdimen\epsftsize    
\newdimen\epsfrsize    
\newdimen\epsftmp      
\newdimen\pspoints     
\def\epsfbox#1{
   \ifvoid\epsfvertlab%
   \else\epsfvlabsize=\ht\epsfvertlab \advance\epsfvlabsize by \dp\epsfvertlab\fi%
   \leavevmode\global\def\epsfllx{72}\global\def\epsflly{72}%
   \global\def\epsfurx{540}\global\def\epsfury{720}%
   \def\lbracket{[}\def\testit{#1}\ifx\testit\lbracket
   \let\next=\epsfgetlitbb\else\let\next=\epsfnormal\fi\next{#1}}%
\def\epsfgetlitbb#1#2 #3 #4 #5]#6{\epsfgrab #2 #3 #4 #5 .\\%
   \epsfsetgraph{#6}}%
\def\epsfnormal#1{\epsfgetbb{#1}\epsfsetgraph{#1}}%
\def\epsfgetbb#1{%
\openin\epsffilein=#1
\ifeof\epsffilein\errmessage{I couldn't open #1, will ignore it}\else
   {\epsffileoktrue \chardef\other=12
    \def\do##1{\catcode`##1=\other}\dospecials \catcode`\ =10
    \loop
       \read\epsffilein to \epsffileline
       \ifeof\epsffilein\epsffileokfalse\else
          \expandafter\epsfaux\epsffileline:. \\%
       \fi
   \ifepsffileok\repeat
   \ifepsfbbfound\else
    \ifepsfverbose\message{No bounding box comment in #1; using defaults}\fi\fi
   }\closein\epsffilein\fi}%
\def\epsfsetgraph#1{%
   \epsfrsize=\epsfury\pspoints
   \advance\epsfrsize by-\epsflly\pspoints
   \epsftsize=\epsfurx\pspoints
   \advance\epsftsize by-\epsfllx\pspoints
   \epsfxsize\epsfsize\epsftsize\epsfrsize
   \ifnum\epsfxsize=0 \ifnum\epsfysize=0
      \epsfxsize=\epsftsize \epsfysize=\epsfrsize
     \else\epsftmp=\epsftsize \divide\epsftmp\epsfrsize
       \epsfxsize=\epsfysize \multiply\epsfxsize\epsftmp
       \multiply\epsftmp\epsfrsize \advance\epsftsize-\epsftmp
       \epsftmp=\epsfysize
       \loop \advance\epsftsize\epsftsize \divide\epsftmp 2
       \ifnum\epsftmp>0
          \ifnum\epsftsize<\epsfrsize\else
             \advance\epsftsize-\epsfrsize \advance\epsfxsize\epsftmp \fi
       \repeat
     \fi
   \else\epsftmp=\epsfrsize \divide\epsftmp\epsftsize
     \epsfysize=\epsfxsize \multiply\epsfysize\epsftmp   
     \multiply\epsftmp\epsftsize \advance\epsfrsize-\epsftmp
     \epsftmp=\epsfxsize
     \loop \advance\epsfrsize\epsfrsize \divide\epsftmp 2
     \ifnum\epsftmp>0
        \ifnum\epsfrsize<\epsftsize\else
           \advance\epsfrsize-\epsftsize \advance\epsfysize\epsftmp \fi
     \repeat     
   \fi
   \ifepsfverbose\message{#1: width=\the\epsfxsize, height=\the\epsfysize}\fi
   \epsftmp=10\epsfxsize \divide\epsftmp\pspoints
   \epsfsetlab%
   \ifincfigs%
     \vbox to\epsfysize{\vfil\hbox to\epsfxsize{%
        \includegraphics{#1}%
        \epsfsetover\hfil}}%
   \else%
     \epsfsetover%
     \vbox to\epsfysize{\hrule\vss\hbox to\epsfxsize{\vrule height
                        \epsfysize\hfil\vrule}\vss\hrule}%
   \fi%
\epsfxsize=0pt\epsfysize=0pt}%

{\catcode`\%=12 \global\let\epsfpercent=
\long\def\epsfaux#1#2:#3\\{\ifx#1\epsfpercent
   \def\testit{#2}\ifx\testit\epsfbblit
      \epsfgrab #3 . . . \\%
      \epsffileokfalse
      \global\epsfbbfoundtrue
   \fi\else\ifx#1\par\else\epsffileokfalse\fi\fi}%
\def\epsfgrab #1 #2 #3 #4 #5\\{%
   \global\def\epsfllx{#1}\ifx\epsfllx\empty
      \epsfgrab #2 #3 #4 #5 .\\\else
   \global\def\epsflly{#2}%
   \global\def\epsfurx{#3}\global\def\epsfury{#4}\fi}%
\def\epsfsize#1#2{\epsfxsize}
\def\ifspace{\ifcat\issp.\else~\fi}
\def\tspace{\futurelet\issp\ifspace}
\def\a{({\it a\kern 1pt})\tspace}
\def\b{({\it b\kern 1pt})\tspace}
\def\c{({\it c\kern 1pt})\tspace}
\def\d{({\it d\kern 1pt})\tspace}
\def\e{({\it e\kern 1pt})\tspace}
\def\f{({\it f\kern 1pt})\tspace}
\def\g{({\it g\kern 1pt})\tspace}
\def\h{({\it h\kern 1pt})\tspace}
\def\i{({\it i\kern 1pt})\tspace}
\def\j{({\it j\kern 1pt})\tspace}
\def\abc#1{({\it #1\kern 1pt})\tspace}

\newcount\ndots
\def\drawline#1#2{\raise 2.5pt\vbox{\hrule width #1pt height #2pt}}

\def\trian{\raise 1.25pt\hbox{$\scriptscriptstyle\triangle$}\nobreak\ }
\def\solidtrian{\raise 1.25pt
\hbox to 3bp{
\hfill}\nobreak\ }
\def\dsolidtrian{\raise 1.25pt
\hbox to 3bp{
\hfill}\nobreak\ }
\def\soliddiamond{\raise 1.25pt
\hbox to 4bp{
\hfill}\nobreak\ }

\def\square{${\vcenter{\hrule height .4pt 
              \hbox{\vrule width .4pt height 3pt \kern 3pt \vrule width .4pt}
          \hrule height .4pt}}$\nobreak\ }

\def\plus{\raise 1.25pt \hbox{$\scriptscriptstyle +$}\nobreak\ }
\def\x{\raise 1.25pt \hbox{$\scriptscriptstyle \times$}\nobreak\ }
\def\legendtable#1{\vbox{\baselineskip=10pt\tabskip=0pt\let\\=\cr\halign{\hfil##\hskip 3pt&##\hfil\cr#1\crcr}}}
\def\lllegend#1 #2 #3{\figlab {#1} {#2} {\legendtable{#3}}}
\def\lrlegend#1 #2 #3{\figlab {#1} {#2} {\llap{\legendtable{#3}}}}
\def\ullegend#1 #2 #3{\figlab {#1} {#2} {\vtop{\hrule height 0pt\legendtable{#3}}}}
\def\urlegend#1 #2 #3{\figlab {#1} {#2} {\llap{\vtop{\hrule height 0pt\legendtable{#3}}}}}

\newdimen\xorigon
\newdimen\yorigon
\newdimen\scaleval
\newdimen\scaleorigon

\def\setxscale#1 #2 #3 #4 #5 {%
    \xorigon=#1\yorigon=#3%
    \scaleval=#2\advance\scaleval by -\xorigon%
    \tempdimen=#5 pt\advance\tempdimen by -#4pt%
    \divide\tempdimen by 1000%
    \divide\scaleval by \tempdimen%
    \scaleorigon=-#4pt\divide\scaleorigon by 1000%
    \multiply\scaleorigon by \scaleval}
\def\xtickup#1 #2{\tempdimen=#1pt\divide\tempdimen by 1000%
    \multiply\tempdimen by \scaleval\advance\tempdimen by \scaleorigon%
    \advance\tempdimen by \xorigon%
    \figlab {\tempdimen} {\yorigon} {\vbox {\hbox to 0pt{\hss #2\hss}%
        \baselineskip=8pt\lineskiplimit=-5pt%
        \hbox to 0pt{\hss \vrule height 3pt\hss}}}}
\def\xtickdown#1 #2{\tempdimen=#1pt\divide\tempdimen by 1000%
    \multiply\tempdimen by \scaleval\advance\tempdimen by \scaleorigon%
    \advance\tempdimen by \xorigon%
    \figlab {\tempdimen} {\yorigon} {\vbox to 0pt {\hbox to 0pt{\hss \vrule height 3pt\hss}%
        \nointerlineskip\vskip 3pt%
        \hbox to 0pt{\hss #2\hss}\vss}}}
\def\nofig#1#2{\leavevmode{\vbox {\hrule \hbox to #1{\vrule height #2 \hfill \vrule} \hrule}} }

\begin{document}
\maketitle
\begin{abstract}
An adjoint-based optimization is applied to study the thrust performance of  a pitching-rolling ellipsoidal plate in a uniform stream at Reynolds number 100. To achieve the highest thrust, the optimal kinematics of pitching-rolling motion is sought in
a large control space including the pitching amplitude, the rolling amplitude, and the phase delay between the pitching and rolling motion.
A continuous adjoint approach with boundary motion being handled by non-cylindrical calculous is developed as a computationally efficient optimization algorithm to deal with the large control space with morphing domain.
The comparison between the optimal motion and other reference motions shows an significant improvement of thrust from the increase of rolling amplitude and an optimal phase delay of $122.6^\circ$  between the pitching and the rolling motion. The combination of these two factors impacts the overall thrust performance through their strong effects on the angle of attack,  circulation, and the pressure distribution on the plate.
Further wake structure analysis suggests that the optimal control improves its propulsive performance by generating a stronger leading-edge vortex (LEV) and straightening the wake deflection.
\end{abstract}

\section{Nomenclature}
{\renewcommand\arraystretch{1.0}
\noindent\begin{longtable*}{@{}l @{\quad=\quad} l@{}}
$\mathbf{q}$ & primary variable \\
$\mathbf{u}$ & velocity \\
$p$  & pressure \\
$\rho$ & density \\
$t$ &  time\\
$x$ &  Cartesian coordinate\\
$\Omega$   & fluid domain \\
$\mathcal S$ & solid boundary \\
$\Gamma_\infty$ & far-field boundary \\
$n$ & normal direction \\
$\mathbf S$ & solid boundary location \\
$\mathbf V$ &  velocity at the solid boundary\\
$\mathbf U$ &  velocity at the far-field\\
$\nu$ & kinematic viscosity \\
$\mathbf{Z}$ & derivative of boundary point location w.r.t control parameters in a Lagrangian framework\\
$\mathcal N({\mathbf q})$ &  operator for Navier-Stokes equations\\
$\phi$ &  control parameters\\
$\mathcal J $ &  objective function\\
$T$ &  flapping period\\
$A$ &  wing area\\
$\sigma_{ij}$ &  stress tensor\\
$\delta_{ij}$ &  Kroneckers delta\\
$\mathcal N^\prime({\mathbf q})\mathbf q^\prime$ &  operator for linearized Navier-Stokes equations\\
$\mathbf{q}^\ast$ &  adjoint variable\\
$\mathbf{u}^\ast$ &  adjoint velocity\\
$p^\ast$ &  adjoint pressure\\
$\mathcal N^\ast(\mathbf q){\mathbf q^\ast}$ & operator for the adjoint equations \\
$g$ & gradient of the objective function with respect to controls parameters \\
$l,c,h$ &  wing span length, mid-chord length, and thickness\\
$\theta_x,\theta_{z^\prime}$ &  rolling and pitching angles\\
$a_x,a_z$ &  amplitudes of rolling and pitching motion \\
$\varphi_z$ &  phase difference between the pitching and rolling motion\\
$f$ &  flapping frequency\\1
$\Delta x, \Delta y, \Delta z$ &  minimum Cartesian grid sizes in  the associated directions \\
$Re$ &  Reynolds number\\
$St$ &  Strouhal number \\
$R$ &  rotational radius\\
$R_{avg}$ &  average rotational radius\\
$\alpha$ & angle of attack \\
$C_T$ & thrust coefficient  \\
$\bar{C_T}$ &  cycle-averaged thrust coefficient \\
$N_{iter}$ &  number of iterations during optimization\\
$\omega_x, \omega_{z^\prime}$ & streamwise and spanwise vorticity \\
$x^\ast, y^\ast$ &  LEV position from leading edge and above wing surface \\
$Q$ &  Q-criterion\\
$\Gamma$ & LEV circulation  \\
$R_1, R_2, R_3$ &  vortex rings\\

\multicolumn{2}{@{}l}{Subscripts}\\

$\dot{()}$ & derivative in a Lagrangian framework\\
$()^\prime$ & derivative in a Eulerian framework\\
$()^\ast$ & adjoint variables or operators
\end{longtable*}}

\section{Introduction}

\lettrine{T}{he} mechanism of flapping motion provides an energy-efficient way for bio-inspired propulsion and is the most common way that has been adopted by flying and swimming animals, such as insects, birds, and fishes. In comparison with conventional man-made designs, flapping propulsors used by natural flyers/swimmers show many attractive characteristics in speed, maneuverability, and high energy efficiency in the low Reynolds number regime \cite{mueller2001overview}.

The flapping motion generally consists of a rolling motion about a fixed joint and a pitching motion with respect to the spanwise axis. Many previous studies applied pitching and/or heaving motion to study the flapping plates through experimental measurements \cite{anderson1998oscillating,koochesfahani1989vortical} or computational simulations \cite{Dong:2006,YangWeiZhao:2010}. Only a few studies can be found on the bio-inspired pitching-rolling propulsors and most of them focused on the investigation of propulsive performance over a range of pitching amplitude and rolling amplitude \cite{bandyopadhyay2012relationship,techet2008propulsive}.
In their investigation of pitching-rolling motion \cite{li2016three}, Li et al. have found a
unique double-C-shaped vortex structure in the wake, which is different from vortex structures generated by pitching-heaving panels or foils.
The finding suggests
that pitching-rolling motion is a more realistic kinematic model for the aero/hydrodynamic investigation of insect/bird wings and fish pectoral fins.
In the same study,
the phase delay between the pitching and rolling motion is suggested playing a critical role in  thrust production. Although careful numerical simulations/experimental measurements have been designed to understand the bio-inspired pitching-rolling motion in aforementioned studies, the large parametric space involved here makes any comprehensive parametric study of flow physics or direct optimization too expensive to be feasible when efforts being extended to cover the full control space.

To achieve an understanding behind a large number of control parameters, one often chooses to reduce the complexity of physical model and/or the size of parametric space.
Based on a quasi-steady model with 11 control parameters, Berman \& Wang were able to use a hybrid algorithm of genetic method and simplex method to minimize the power consumption of insect flights \cite{berman2007energy}.
Ghommem {\it et al.} used unsteady vortex-lattice method and a deterministic global optimization algorithm for the optimization of flapping wings in forward flight with active morphing, where only 4 to 8 parameters were considered \cite{ghommem2012global}.
Milano \& Gharib applied genetic algorithm in an experimental setting to maximize the average lift from a flapping flat plate by limiting the number of control parameters to only 4 \cite{milano2005uncovering}.
Trizila {\it et al.}  used a combined approach with numerical simulation and surrogate modeling to explore a three-parameter design space for a three-dimensional plate in hovering motion \cite{trizila2011low}. Building a map of an entire parameter space is useful for some problems but may not be necessary for others. On the other hand, the computation cost was still high even with a surrogate model and prevents the study of more parameters, and the accuracy was limited by the chosen surrogate model.

Different from the above approaches, an adjoint-based method is capable of obtaining the gradient information simultaneously for an arbitrary number of input parameters by one single computation in adjoint space. Consequently, the total computational cost to obtain the sensitivity of a cost function to all control parameters is independent of the number of control parameters. With the sensitivity obtained, gradient-based optimization algorithms can be used to find the local optimal solution efficiently. Thus, it makes an adjoint-based method suitable for the sensitivity analysis and optimization of problems with a large input space but a small output space. There are two types of adjoint approaches: continuous approach \cite{Bewley:2001, Wei:2006} and discrete approach \cite{jones2013adjoint,lee2011non}. In the current work, we take the continuous approach, considering its advantage of simplicity and clarity in the governing equation over the discrete approach \cite{nadarajah2000comparison,Collis:2001}. Jameson \cite{jameson2003aerodynamic} used a continuous adjoint approach to optimize aerodynamic shape designs in both inviscid and viscous compressible steady flow in a fixed-domain setup. Later, Nadarajah \& Jameson \cite{jones2013adjoint} used continuous and discrete adjoint approaches on the shape optimization of two-dimensional oscillating airfoils where moving boundary was presented. Using a time-varying mapping function to transfer the physical domain with moving boundary to a computational domain with fixed boundary, the traditional adjoint-based method for a fixed domain may be adopted to apply directly on morphing-domain problems \cite{jones2013adjoint}. However, the complexity of formulation from using a time-varying mapping function increases dramatically with the complexity of the problem to a point such that it becomes intractable in dealing with a three-dimensional (or even two-dimensional) problem described by Navier-Stokes equations.
Recently, Xu and Wei applied non-cylindrical calculus \cite{Moubachir:2006, Protas:2008}, instead of a time-varying mapping function, to derive adjoint equations  for a morphing domain directly in its physical space \cite{xu2016using}, where the derivation process and the final formulation were simple enough for the study of three-dimensional morphing-domain problems with flows described by Navier-Stokes equations.

In this paper, the same adjoint-based approach as in \cite{xu2016using} allows us to study the complex flow physics behind a three-dimensional pitching-rolling plate for its impact on thrust performance in a forward flight.
In order to achieve an optimal cycle-averaged thrust production, three control parameters are considered in the optimization:  the pitching amplitude, the rolling amplitude, and the phase delay between the pitching and rolling motion. The rest of the paper begins by describing the theoretical derivation and numerical algorithms in section \S~\ref{sec:methodology}. Then, the flow physics behind the propulsive performance of a pitching-rolling plate is investigated using the adjoint-based optimization approach in section \S~\ref{sec:results}. Finally, the conclusions are summarized in section \S~\ref{sec:conclusion}.

\section{Methodology} \label{sec:methodology}
Non-cylindrical calculus is applied to formulate the adjoint equation system in a morphing domain for the current study. The basics of theoretical derivation and numerical implementation of the approach are provided here, while more details may be referred to earlier works \cite{XuWei:2013,xu2014continuous,xu2015adjoint,xu2016using}.

\subsection {Governing equation and cost function}
The flow is described by the incompressible Navier-Stokes equations, where all the variables are non-dimensionalized accordingly by
the spanwise wing length, incoming velocity, and fluid density, as
\begin{equation}\label{eqn:NS1}
\begin{aligned}
\mathcal N({\mathbf q}) &= \mathbf{0} \qquad \mathrm{in}\;\; \Omega,\\
\mathbf u &= \mathbf V \qquad \mathrm{on} \;\; \mathcal S,\\
\mathbf u &= \mathbf U \qquad \mathrm{on} \;\; \Gamma_\infty,\\
\frac{\partial p} {\partial n} &= 0\qquad \mathrm{on} \;\; \Gamma_\infty,
\end{aligned}
\end{equation}
where $\Omega$ represents the fluid domain, $\mathcal S$ and $\Gamma_\infty$ denote the solid boundary and the far-field boundary respectively, $n$ is the unit normal vector at  $\Gamma_\infty$,  the flow variable $\mathbf q=[p\; \mathbf u]^T$ includes the pressure $p$ and the velocity vector $\mathbf u$,  $\mathbf V$ and $\mathbf U$ are velocity vectors at corresponding boundaries. The Navier-Stokes operator $\mathcal N$ is
 \begin{equation}
\mathcal N(\mathbf q) = \left[\begin{aligned}
   & \frac{\partial u_j}{\partial x_j} \\
   \frac{\partial u_i}{\partial t} + \frac{\partial u_j u_i}{\partial x_j} &-\nu \frac{\partial^2 u_i}{\partial x^2_j} +\frac{1}{\rho}\frac{\partial p}{\partial x_i}  \\
      \end{aligned}\right],
\end{equation}
with the Einstein summation convention being implied for repeated indices, $\nu$ denoting the nondimensionalized kinematic viscosity, and $\rho$ being the density which equals 1 due to aforementioned non-dimensionalization for incompressible flow. The wing motion is prescribed by flapping motion with a set of control parameters $\phi$, therefore solid boundary location and velocity are functions of control parameters and can be expressed as: $S_i=S_i(\phi, t)$ and $V_i=V_i(\phi, t)$.

To optimize the thrust performance, the negative of thrust coefficient is chosen to be the cost function to form a minimization problem:
\begin{equation}
\begin{aligned}
\mathcal J = -\frac {1}{T D_0} \int_0^T \int_{\mathcal S} {\boldsymbol \sigma_1} \cdot {\mathbf n} \text{d}s \text{d}t,\\
\end{aligned}
\end{equation}
where $T$ is the flapping period, $D_0=  U^2 A/2$ with $A$ being the wing area, and
\begin{equation}
\sigma_{ij} = -\frac{p}{\rho} \delta_{ij} + \nu  \left(\frac{{\partial {{u}_i}}}{{\partial {x_j}}} + \frac{{\partial {{u}_j}}}{{\partial {x_i}}} \right),
\end{equation}
with $\delta_{ij}$ Kronecker’s delta, so $\mathbf {\boldsymbol \sigma_1} = \sigma_{1j}$ represents the stress contributing to the thrust force along $x$ direction.

\subsection {Linearized perturbation equation and perturbed cost function}
It has been demonstrated that non-cylindrical calculus has great advantages in efficiency and simplicity in the derivation of an adjoint equation in continuous form for moving boundary problems \cite{Moubachir:2006, Protas:2008, xu2016using}. Following the same derivation, we can easily derive the linearized perturbation equation for the Navier-Stokes equations using the shape derivative:
\begin{equation}\label{eqn:linear_eqn1}
\begin{aligned}
\mathcal N^\prime(\mathbf q){\mathbf q^\prime} &= \mathbf{0} \qquad \mathrm{in}\;\; \Omega,\\
{\mathbf u^\prime} &= {\dot {\mathbf V}} - {\mathbf Z} \cdot \nabla {\mathbf u} \;\; \mathrm{on} \;\; \mathcal S, \\
{\mathbf u^\prime} &= 0 \qquad  \mathrm{on} \;\; \Gamma_\infty,\\
\frac{\partial p^\prime} {\partial n}  &= 0 \qquad  \mathrm{on} \;\; \Gamma_\infty,\\
\end{aligned}
\end{equation}
where $\dot{()}$ indicates the derivative w.r.t control parameters in a Lagrangian framework, and $()^\prime$ indicates the  derivative in an Eulerian framework, ${\mathbf Z} = \dot {\mathbf S}$ with $\mathbf S$ denoting the location of boundary points, and
\begin{equation}\label{eqn:linear_eqn2}
\mathcal N^\prime(\mathbf q)^\prime = \left[\begin{aligned}
   & \frac{\partial u_j^\prime}{\partial x_j} \\
   \frac{\partial u_i^\prime}{\partial t} + \frac{\partial u_j^\prime u_i}{\partial x_j} +& \frac{\partial u_j u_i^\prime}{\partial x_j}-\nu \frac{\partial^2 u_i^\prime}{\partial x^2_j} +\frac{1}{\rho}\frac{\partial p^\prime}{\partial x_i}  \\
      \end{aligned}\right].
\end{equation}

Following \cite{xu2016using}, the functional derivative of the cost function subject to the same perturbation is given by
\begin{equation}
\begin{aligned}
\mathcal J^\prime = -\frac {1}{T D_0} \int_0^T \int_{\mathcal S} \left ( {\boldsymbol \sigma_1}^\prime \cdot {\mathbf n} + (\nabla \cdot {\boldsymbol \sigma_1}) (\mathbf Z \cdot \mathbf n) \right)  \text{d}s \text{d}t.\\
\end{aligned}
\end{equation}

\subsection {Adjoint equation and gradient calculation} \label{sec:adjoint_equation}
Adjoint variables $\mathbf q=[p^\ast\; \mathbf u^\ast]^T$ are introduced as Lagrange multipliers to impose the flow equations, so that we obtain the derivative of the enhanced cost function,
\begin{equation}
\begin{aligned}
\mathcal J^\prime =  -\frac {1}{T D_0} \left (\int_0^T \int_{\mathcal S} \left ( {\boldsymbol \sigma_1}^\prime \cdot {\mathbf n} + (\nabla \cdot {\boldsymbol \sigma_1}) (\mathbf Z \cdot \mathbf n) \right)  \text{d}s \text{d}t + \int_0^T \int_{\Omega} {\mathbf q}^\ast \cdot \mathcal N^\prime(\mathbf q) \mathbf q^\prime \text{d}\Omega \text{d}t \right).\\
\end{aligned}
\end{equation}

Using integration by parts to group and separate the perturbation terms, we then have
\begin{equation}
\begin{aligned}
\mathcal J^\prime =  -\frac {1}{T D_0}\left( b -  \int_0^T \int_{\Omega} {\mathbf q}^\prime \cdot \mathcal N^\ast(\mathbf q) \mathbf q^\ast \text{d}\Omega \text{d}t\right),\\
\end{aligned}
\end{equation}
where
\begin{equation}\label{eqn:NS-adj}
\mathcal N^\ast(\mathbf q){\mathbf q^\ast} =\left[\begin{aligned}
    &\frac{1}{\rho} \frac{\partial u_j^\ast}{\partial x_j} \\
   \frac{\partial u_i^\ast}{\partial t} + u_j\left(\frac{\partial u_i^\ast}{\partial x_j} + \frac{\partial u_j^\ast}{\partial x_i} \right)+&\nu \frac{\partial^2 u_i^\ast}{\partial x^2_j} +\frac{\partial p^\ast}{\partial x_i}  \\
      \end{aligned}\right].
\end{equation}
The boundary terms from integration by parts are all included in $b$ as
\begin{equation}
\begin{aligned}
b =   &\left. \int_{\Omega} u^\ast_j u^\prime_j \text{d}\Omega \right|_{t=0}^{t=T} + b_{\infty} + \int_0^T \int_{\Omega} (u^\ast_i+\delta_{1i})\sigma^\prime_{ij}n_j \text{d}s\text{d}t\\
&- \int_0^T \int_{\Omega} u_i^\prime(\sigma_{ij}^\ast n_j+u_j^\ast u_j n_i) \text{d}s\text{d}t + \int_0^T \int_{\Omega} Z_{k} \frac{\partial \sigma _{1j}} {\partial x_j} n_k \text{d}s\text{d}t,
\end{aligned}
\end{equation}
where $b_\infty$ includes all far-field terms and
\begin{equation}
\sigma_{ij}^* =  p^\ast \delta_{ij} + \nu  \left(\frac{{\partial {u^*_i}}}{{\partial {x_j}}} + \frac{{\partial {u^*_j}}}{{\partial {x_i}}} \right).
\end{equation}

It is noticed that the above formulation of $\mathcal J^\prime$ can be largely simplified if the following conditions are imposed:
\begin{equation}\label{eqn:ad_eqn1}
\begin{aligned}
\mathcal N^\ast(\mathbf q){\mathbf q^\ast} &= \mathbf{0} \qquad \mathrm{in}\;\; \Omega,\\
{\mathbf u}^\ast &=  -{\boldsymbol \delta_1} \;\; \mathrm{on} \;\; \mathcal S, \\
{\mathbf u}^\ast &= 0 \qquad  \mathrm{on} \;\; \Gamma_\infty,\\
p^\ast   &= 0 \qquad  \mathrm{on} \;\; \Gamma_\infty,\\
\end{aligned}
\end{equation}
where ${\boldsymbol \delta_1} = \delta_{1j}$. When both the flow and adjoint solutions are periodic, the gradient of cost function $\mathcal J$ with respect to controls $\mathbf{\phi}$ is then given by,
\begin{eqnarray}
\begin{aligned}
g_l= \frac {\partial \mathcal J} {\partial \phi_l} = -\frac{1}{TD_0} \int_0^T \int_{\mathcal S} \left [ Z_{k,l} \frac{\partial \sigma _{1j}} {\partial x_j} n_k - \left( {\dot V}_{i,l} - Z_{k,l} \frac {\partial u_i} {\partial x_k} \right) Z^*_i \right ] ds dt,
\end{aligned}
\end{eqnarray}
where
\begin{eqnarray}
\begin{aligned}
{\dot V}_{i,l} = \frac {\partial V_i} {\partial \phi_l},\quad {Z}_{i,l} = \frac {\partial S_i} {\partial \phi_l},\quad
Z^*_i = \sigma _{ij}^* n_j + u_j^* u_j n_i.
\end{aligned}
\end{eqnarray}
The control being updated by the gradient leads to the decrease of the cost function in optimization.

\subsection{Numerical algorithm} \label{sec:Numerical}
For both the forward (flow) simulation and the backward (adjoint) simulation, we used immersed boundary method \cite{mittal2008versatile} to treat moving boundaries. This immersed-boundary-method-based simulations have been widely used to simulate the bio-inspired flapping locomotion \cite{li2017wing,liu2016vortex,li2015effects}. A staggered Cartesian mesh with local refinement through stretching functions is chosen for the benefit of both computational efficiency and numerical stability. We used the central difference for spatial discretization, second-order Adams-Bashforth/Crank-Nicolson scheme for time advancement and projection method to keep incompressible constrains \cite{YangWeiZhao:2010,xu2016embedded,xuTheis:2014}, where Possion equation is solved by FFT and a generalized cyclic reduction algorithm \cite{yangTheis:2005}. The time step is limited by the CFL constraint. With the similarity shown in the form of adjoint equations and flow equations,  similar numerical algorithms are implemented to solve the adjoint equations backward in time for about the same computational cost. To reduce the data storage, the flow solutions required in the adjoint simulation are saved every other time step and in single precision. 

The gradient of the cost function w.r.t. all the control parameters can be calculated after both the flow and adjoint equations are solved once. In the current study, we use the incoming velocity as the initial condition for the flow simulation and zero velocity as the initial condition for the adjoint simulation. Both the flow and adjoint simulations reach the periodic state after 3 flapping periods, based on the monitoring of the instantaneous force or gradient. Therefore, we ran the forward flow simulations for a total of 8 periods and the backward adjoint simulation for 5 periods, and use the flow and adjoint solutions at $3T<t<5T$ for the final gradient calculation.

In the main iteration to update the gradient, the Polak-Ribiere variant of the conjugate gradient method is used, which has been tested and used before in many of the related work \cite{Wei:2006,XuWei:2013,xu2014continuous,xu2015adjoint,xu2016using}; within each main iteration, COBYLA scheme (constrained optimization by linear approximations) \cite{powell1994direct} in NLopt software package \cite{johnson2014nlopt} is used to determine the optimal step size within the constrained range along each direction and this process requires about 6-9 subiterations of flow simulation.

\section{Thrust Study of a Three-Dimensional Pitching-Rolling Plate} \label{sec:results}

In this section, we first use the adjoint-based optimization approach to investigate the thrust production of a rigid pitching-rolling  plate. The control parameters include the pitching amplitude, the rolling amplitude, and the phase delay between the pitching and rolling motion. To reveal the underlying flow physics of the plate undergoing the optimal pitching-rolling motion, three additional reference control cases are further examined and compared with the initial and the optimal control cases.

\subsection {Kinematics and computational setup}
\begin{figure}
\centerline{
\includegraphics[width=0.9\textwidth]{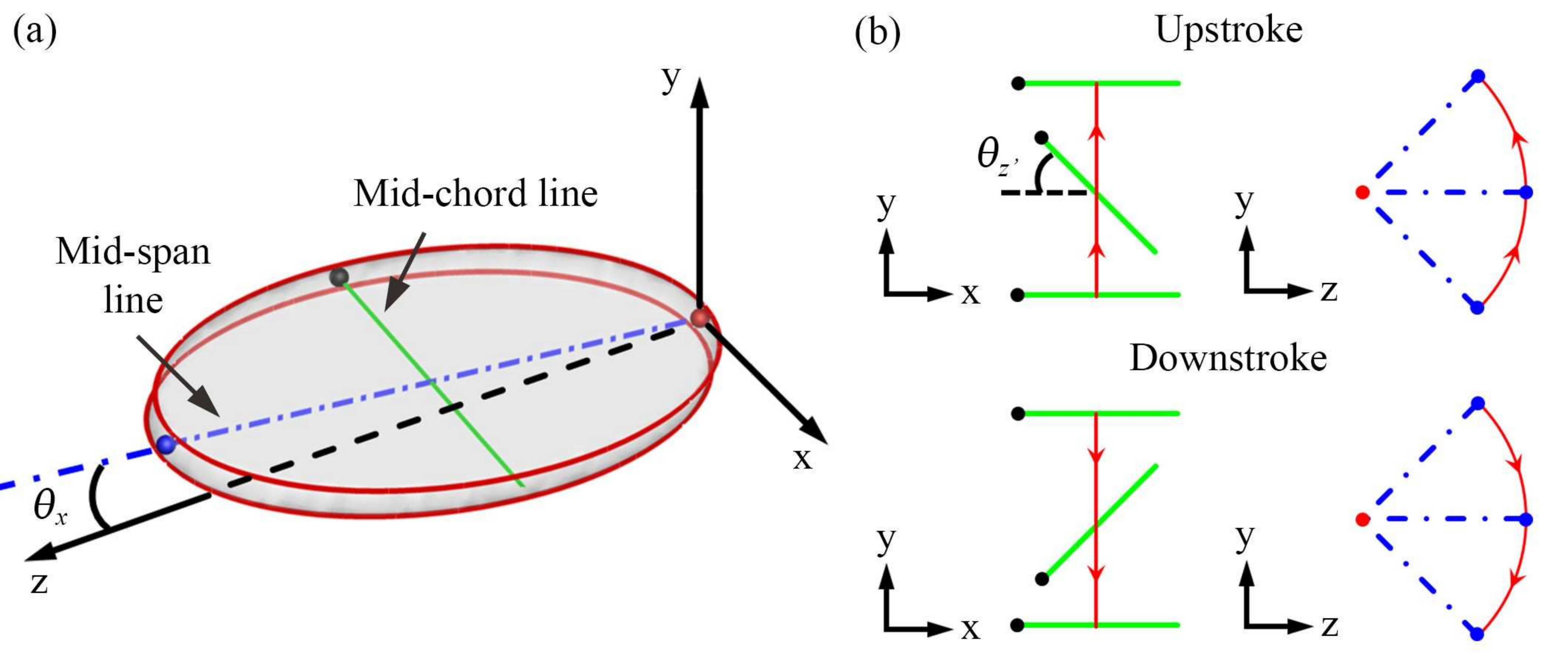}}
\caption{Schematic of the plate kinematics: (a) the perspective view of the three-dimensional plate in global coordinate, and (b) the two-dimensional views of the mid-chord line and mid-span line kinematics during upstroke and downstroke.}
\label{fig:motion}
\end{figure}

\begin{figure}
\centerline{
\includegraphics[width=0.6\textwidth]{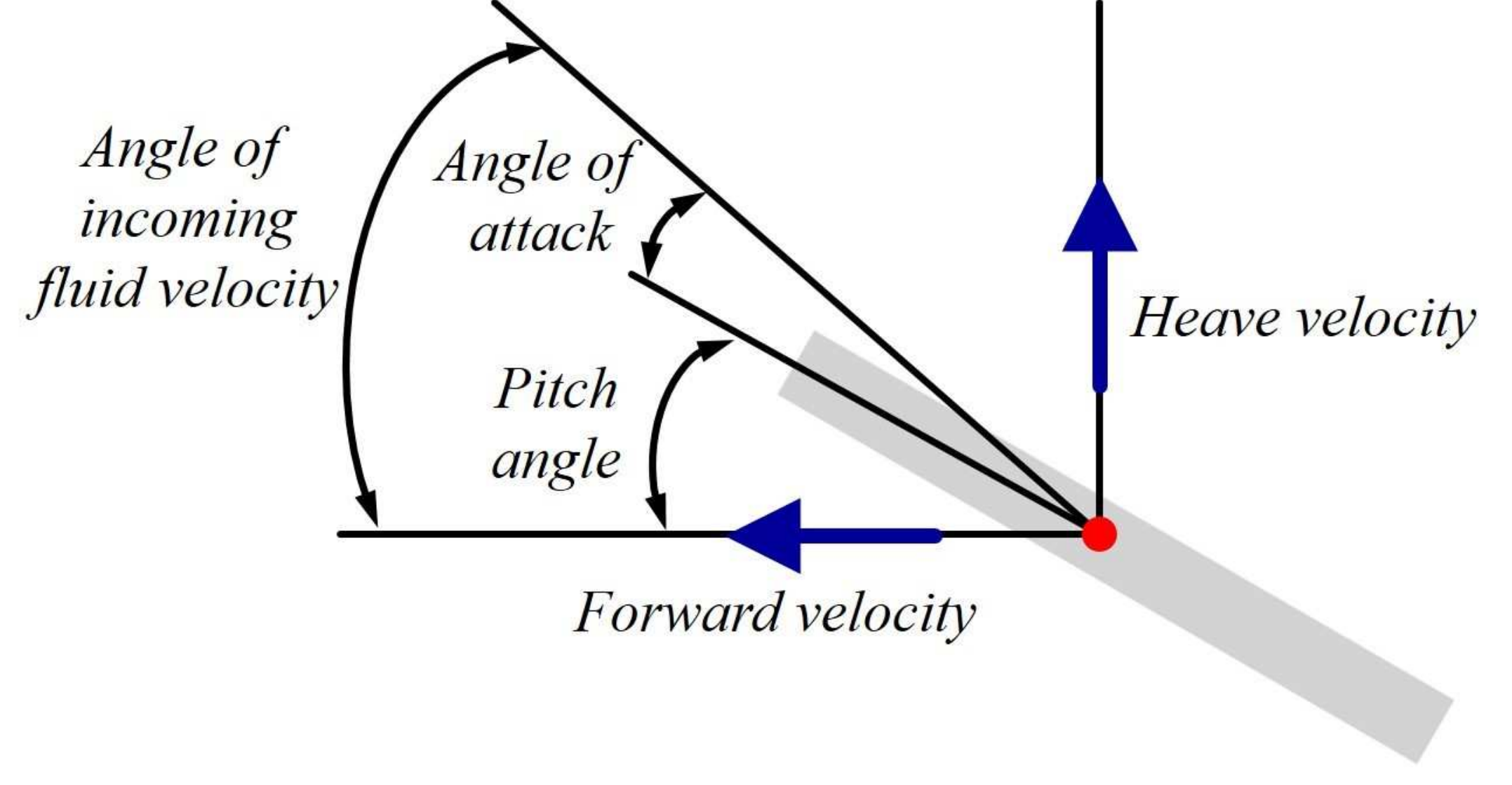}}
\caption{Definition of the angle of attack for a wing chord relative to pitching and heaving motions and incoming fluid velocity.}
\label{fig:AOA}
\end{figure}

An ellipsoidal plate is used in the current study, with the non-dimensional span length of $l=1$ after non-dimensionalization, the mid-chord length $c=0.5$ and thickness $h=0.05$. 
The plate is oriented based on a fixed point at the root, as shown in Figure  \ref{fig:motion}. The rolling motion of the plate is along the global $x$ axis and the pitching motion is respect to its local spanwise $z^\prime$ axis. The pitching and rolling equations are given by
\begin{equation}\label{eqn:motion}
\begin{aligned}
\theta_x &= -a_{x}\sin(2 \pi f t),\\
\theta_{z^\prime} &= -a_{z}\sin(2 \pi f t + \varphi_z),
\end{aligned}
\end{equation}
where $\theta_x$ and $\theta_{z^\prime}$ are the instantaneous rolling and pitching angles, $a_{x}$ and $a_{z}$ are the amplitudes of rolling and pitching motion, $\varphi_z$ is the phase difference between the pitching and rolling motion, and $f=1$ is the flapping frequency.

The Reynolds number is defined as $Re=U_\infty c/\mu =100$, based on the incoming flow velocity ($U_\infty$) and the mid-chord length ($c$). The Strouhal number is defined as $St=2a_xR_{avg}f/U_\infty$  based on the rolling amplitude and the average rotational radius ($R_{avg}=l/2$). During the pitching-rolling motion, the angle of attack profile varies along the span of the three-dimensional flapping plate. At any span location on the plate, the kinematics may be decomposed to two-dimensional heaving and pitching. Figure  \ref{fig:AOA} illustrates the vector diagram of the velocity components. Based on the diagram, the angle of attack can be expressed as
\begin{equation}\label{eqn:AOA}
\begin{aligned}
\alpha(t)=\tan^{-1}\left( \frac{-2\pi f a_x R cos(2 \pi f t)}{U_\infty}\right)-\theta_{z^\prime}(t)
\end{aligned}
\end{equation}

A constant inflow velocity ($U_\infty$) boundary condition is applied on the front wall along the axis and the lateral boundaries. The back wall of the  axis is the outflow boundary condition, allowing the vortices to convect out of this boundary without significant reflections. A homogeneous Neumann boundary condition is used for the pressure at all boundaries.

\subsection {Solid and fluid meshes}

\begin{figure}
\centerline{
\includegraphics[width=0.9\textwidth]{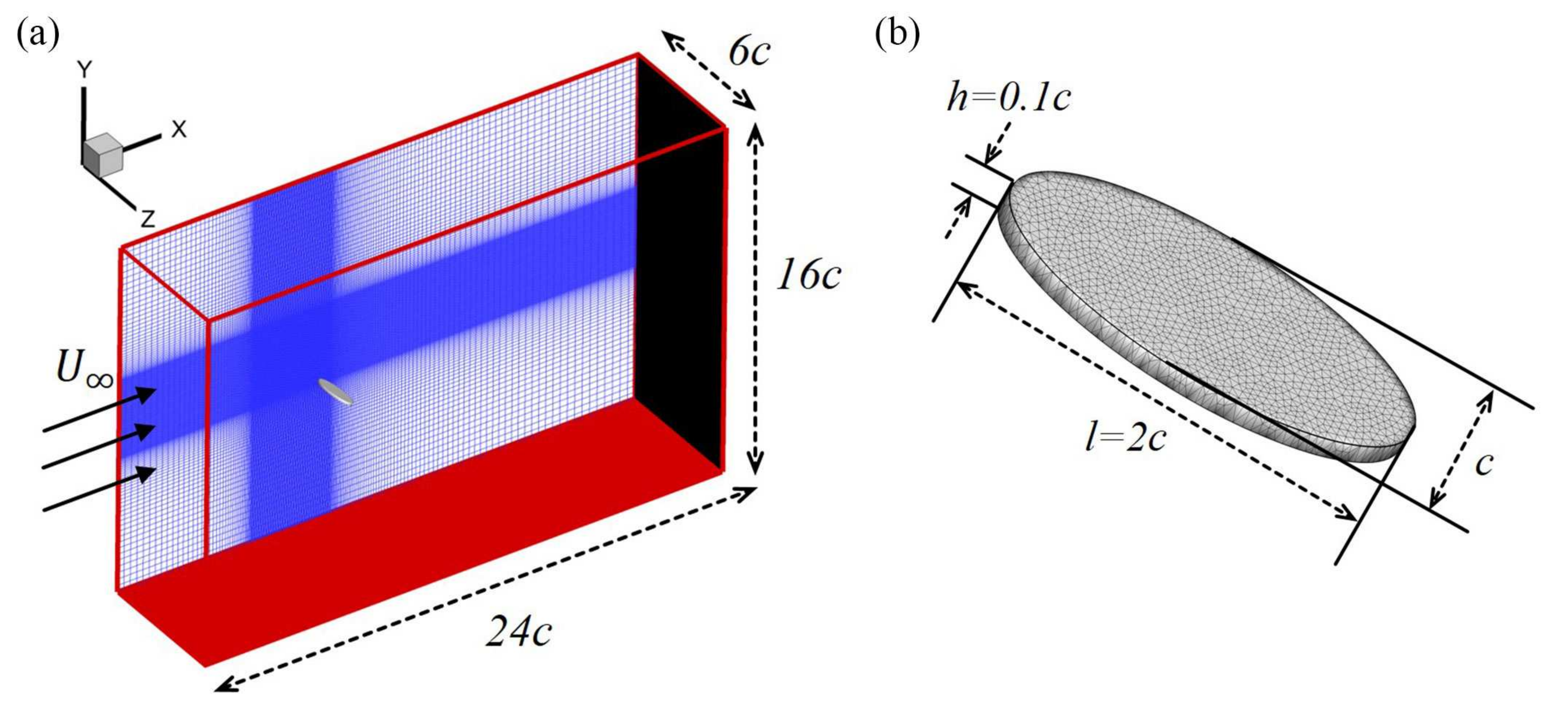}}
\caption{Schematics of (a) the computational domain and Cartesian mesh employed in the current simulation and (b) a typical ellipsoidal plate used in the current study. The surface of the plate is represented by unstructured meshes made of triangular elements.}
\label{fig:3D_Wing}
\end{figure}

As shown in Figure  \ref{fig:3D_Wing}, the surface of the ellipsoidal plate is discretized by 4536 unstructured triangle mesh. A Cartesian mesh, stretched in $x$ and $y$ directions and uniform in $z$ direction, is used for an overall Eulerian description of the combined fluid and solid domain, where uniform grid is adopted in $z$ direction due to the requirement of FFT in Possion solver mentioned previously. The Cartesian grid is refined and clustered uniformly near solid region.

Table \ref{tab:grid} lists 4 different grids used in the current study for grid/domain independence. The coarse mesh provides results qualitatively right but is not sufficient for mesh independence.
The normal grid at $240\times200\times200$ (9.6 million total grid points for fluid) in a domain $24c\times16c\times6c$ with a minimum spatial resolution at $\Delta x = \Delta y = \Delta z = 0.03c$ shows a good balance between computational cost and accuracy, and is the main mesh to provide computational results for analysis and discussion for the rest of the paper. The minimum spatial resolution of the normal grid is the same as the one used in the simulation of a similar study \cite{li2016three}.
For the study of convergency in grid refinement,  the grid size near the solid region is reduced in all directions to have a minimum spatial resolution at $\Delta x = \Delta y = \Delta z = 0.02c$ , and it leads to a grid mesh at $300\times260\times300$ for a total of 23.4 million grid points to keep the same domain size. The time step seize is reduced accordingly in simulation.
For the study of domain independence, the domain size is enlarged by 50\% in all direction to $36c\times24c\times9c$.
To keep the same spatial resolution, the grid mesh is at $300\times240\times300$ for a total of 21.6 million grid points. The thrust coefficient and the norm of the gradient based on the computation using the above 4 different grids are compared in Table \ref{tab:grid} for the same flapping case with the control parameters of $a_x=30^\circ$, $a_z=30^\circ$, and $\varphi_z=90^\circ$.
The grid with enlarged domain shows almost identical results as the normal grid, and the finer mesh grid leads to a difference of the mean thrust for less than 3.5\% and the gradient norm for less than 2.5\%.

\begin{table}
\caption{Result of grid and domain independence study at $a_x=30^\circ$, $a_z=30^\circ$, and $\varphi_z=90^\circ$.}
\begin{center}
\begin{tabular}{cccccc}
 \hline
  Grid & Grid size & Domain size & Minimum grid size &${\bar C}_T$  & $|g|$ \\
  Coarse grid  & $200\times140\times120$ & $24c\times16c\times6c$ & 0.05c & 0.320 & 0.0178 \\
  Nominal grid  & $240\times200\times200$ & $24c\times16c\times6c$  & 0.03c  & 0.286 & 0.0182 \\
  Finer mesh grid &$300\times260\times300$ & $24c\times16c\times6c$& 0.02c  & 0.276 & 0.0186 \\
  Larger domain grid &$300\times240\times300$ & $36c\times24c\times9c$ & 0.03c  & 0.286 & 0.0182 \\
 \hline
\end{tabular}
\end{center}
\label{tab:grid}
\end{table}

%

\subsection{Optimization results}

The control parameters of the pitching and rolling motions include pitching amplitude ($a_x$), rolling amplitude ($a_z$), and the phase delay ($\varphi_z$) between the pitching and the rolling motion. The control,
\begin{equation}
\phi=(a_x,a_z,\varphi_z),
\end{equation}
is optimized to improve the propulsive force, with the parameters being optimized in the range defined in Table \ref{tab:range}.
The maximum values for the amplitudes $a_x$ and $a_z$ are suggested to reasonably represent the kinematics of the bluegill sunfish pectoral fin \cite{lauder2005design}, where it is shown that the propulsive performance of the foils/panels would start to decrease once the pitching amplitude ($a_z$) increases above $45^\circ$. The current upper limit of the rolling amplitude ($a_x=45^\circ$) results in a possible total of $90^\circ$ up-down flapping range. Further increase of the rolling limit may also lead to ground effect between the flapping propulsor and the propelled body wall in engineering applications.
\begin{table}
\caption{The range of control variables for thrust optimization.}
\begin{center}
\begin{tabular}{ccc}
 \hline
  Parameters & Minimum  & Maximum \\
  $a_x$    & $0^\circ$ & $45^\circ$ \\
  $a_z$  & $0^\circ$ & $45^\circ$ \\
  $\varphi_z$  & $-180^\circ$ & $180^\circ$\\
 \hline
\end{tabular}
\end{center}
\label{tab:range}
\end{table}

\begin{figure}
\centerline{
\includegraphics[width=0.85\textwidth]{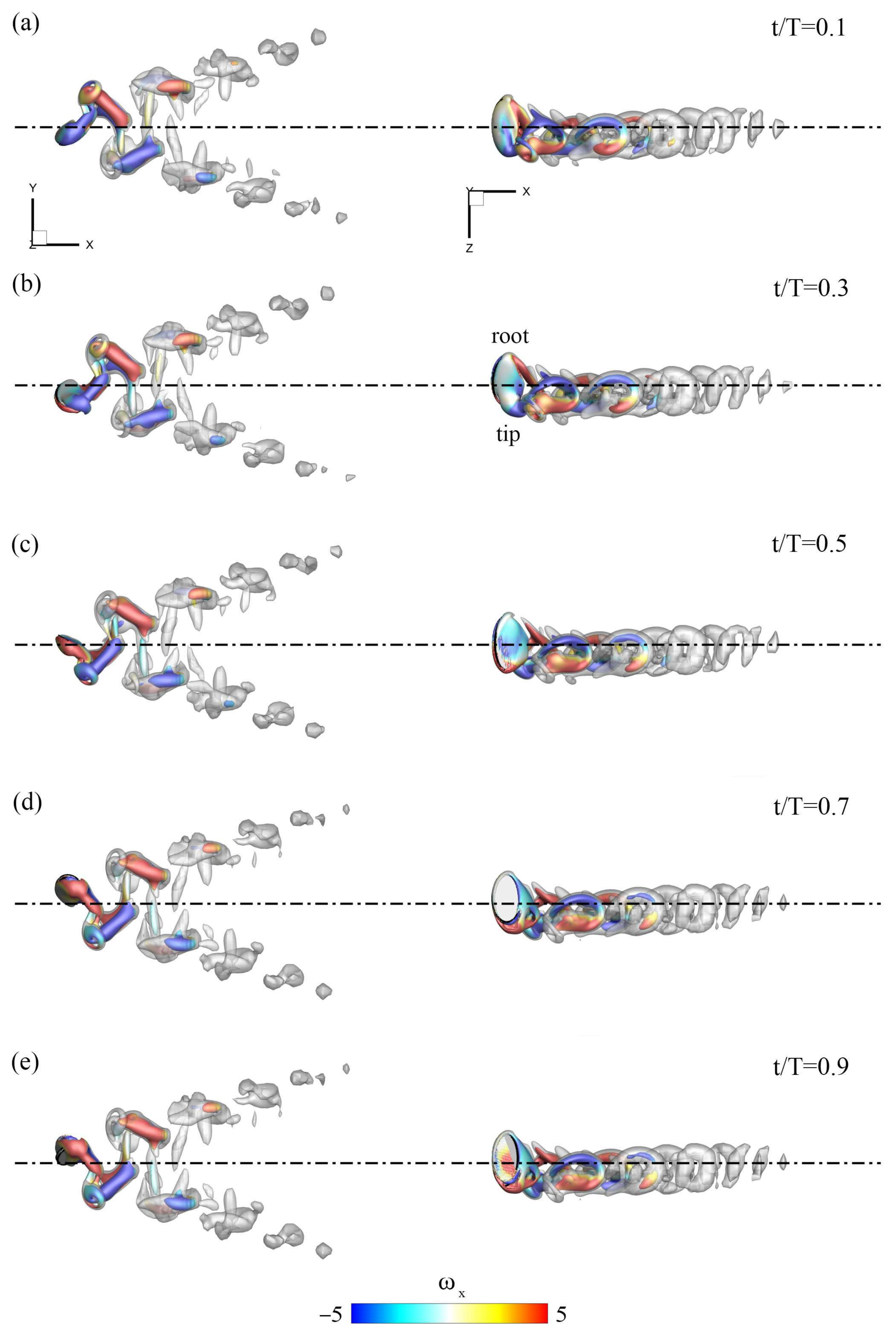}}
\caption{Wake topology of the plate with the initial control at five selected instants: (a) $t/T$=0.1, (b) $t/T$=0.3, (c) $t/T$=0.5, (d) $t/T$=0.7, and (e) $t/T$=0.9. The left and right columns show the side view and top view, respectively. The iso-surface contours are color coded by the streamwise vorticity ($\omega_x$).}
\label{fig:wakeOverPeriod}
\end{figure}

\begin{figure}
\centerline{
\includegraphics[width=1\textwidth]{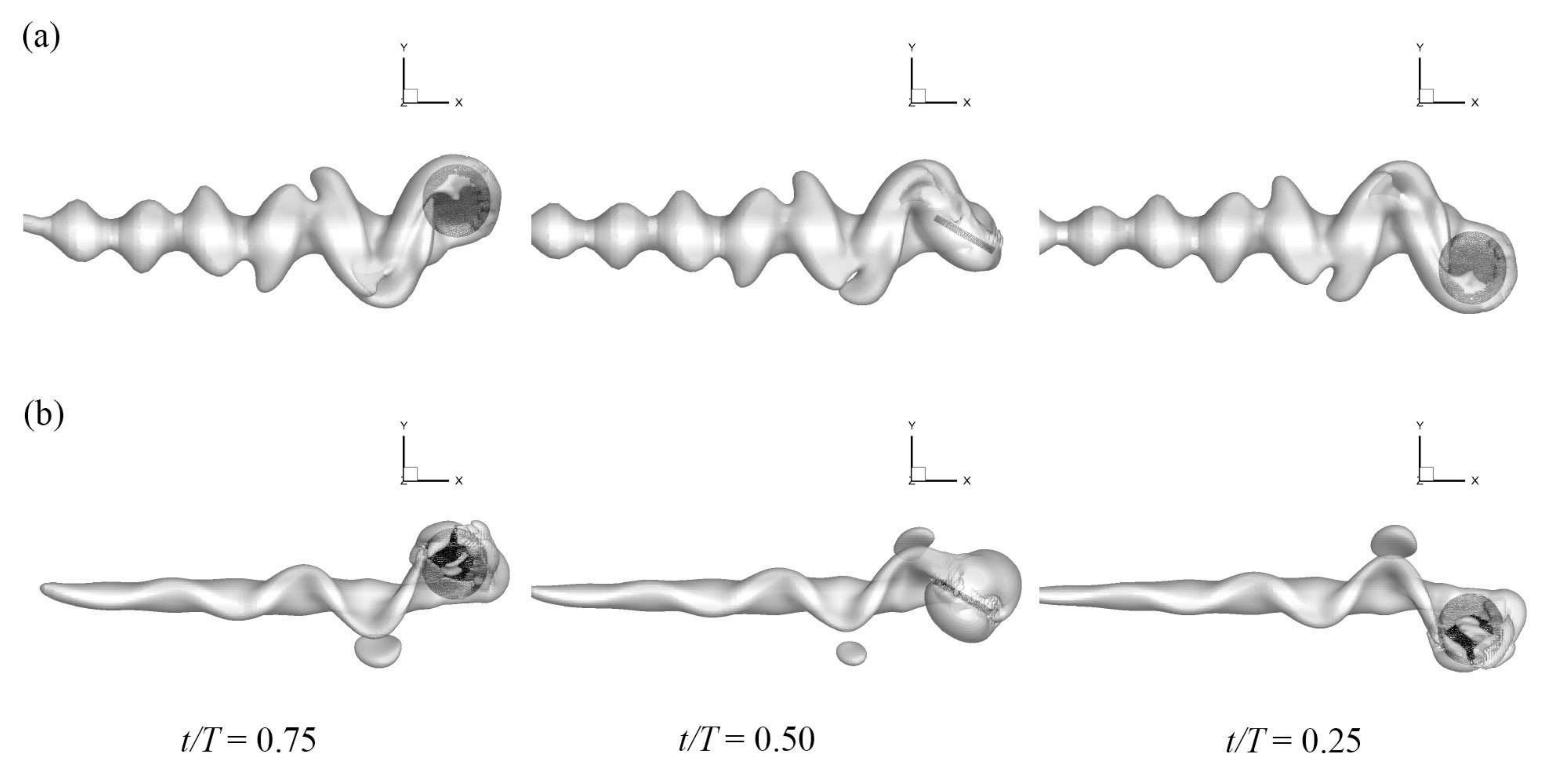}
}
\caption{The iso-surfaces of (a) adjoint velocity magnitude at $|u^\ast|=0.3$ and (b) adjoint pressure magnitude at $|p^\ast|=0.3$ for the pitching-rolling plate with the initial control at $t/T=0.75$, $t/T=0.5$, and $t/T=0.25$.}
\label{fig:adVel}
\end{figure}

Starting with an arbitrary but reasonable initial control: $\phi^0=(30^\circ, 30^\circ, 90^\circ)$, the flow field and adjoint field are each simulated once to establish some initial understanding of the problem.
Figure  \ref{fig:wakeOverPeriod} plots the wake topology of the flapping plate at 5 instants over a period. The shell and core of the vortex structures are visualized by using two layers of Q-criterion \cite{Haller:2005}, $Q = 0.5$ (in grey) and $Q = 5$ (in color), respectively. The vortex cores are color coded by the streamwise vorticity, $\omega_x$. As the plate rolls downward and upward, a pair of vortex rings are produced from each flapping cycle and resulted in a bifurcated wake pattern in the downstream. Figure  \ref{fig:adVel} (a) shows the iso-surfaces of the adjoint velocity magnitude with the initial control. The adjoint field mainly originates from the solid boundary condition, convects upstream by the convective terms, and is gradually damped out by the dissipation term. The adjoint velocity can be interpreted as the sensitivity (gradient) of the cost function with respect to an infinitesimal body force in the momentum equation \cite{wang2012drag}, as it is explained further in details in the Appendix. Figure  \ref{fig:adVel} (b) plots the iso-surfaces of the adjoint pressure magnitude, whose flow pattern is similar to adjoint velocity. It can be interpreted as the sensitivity (gradient) of the cost function with respect to an infinitesimal mass sources and sinks in the continuity equation.

%

%

\begin{table}
\caption{The control parameters and thrust coefficients for the pitching-rolling plates simulated in the current study. }
\begin{center}
\begin{tabular}{ccc}
 \hline
  Case & $\phi$  & ${\bar C}_T$ \\
 \hline
  Initial    & $(30.0^\circ, 30.0^\circ, 90.0^\circ)$ & $ 0.286$ \\
  Optimized  & $(45.0^\circ, 35.9^\circ, 122.6^\circ)$ & $ 2.390$ \\ 
  Reference-0  & $(30.0^\circ, 35.9^\circ, 122.6^\circ)$ & $ 0.580$ \\
  Reference-1  & $(45.0^\circ, 30^\circ, 122.6^\circ)$ & $ 2.284$ \\
  Reference-2  & $(45.0^\circ, 35.9^\circ, 90.0^\circ)$ & $ 1.991$ \\
 \hline
\end{tabular}
\end{center}
\label{tab:drag_3D}
\end{table}

\begin{figure}
\centerline{
\figlab 0.in -0.in {(a)}
\sethlabel {$t/T$}
\setvlabel {$\theta_x$ ($^\circ$)}
\includegraphics[width=0.35\textwidth]{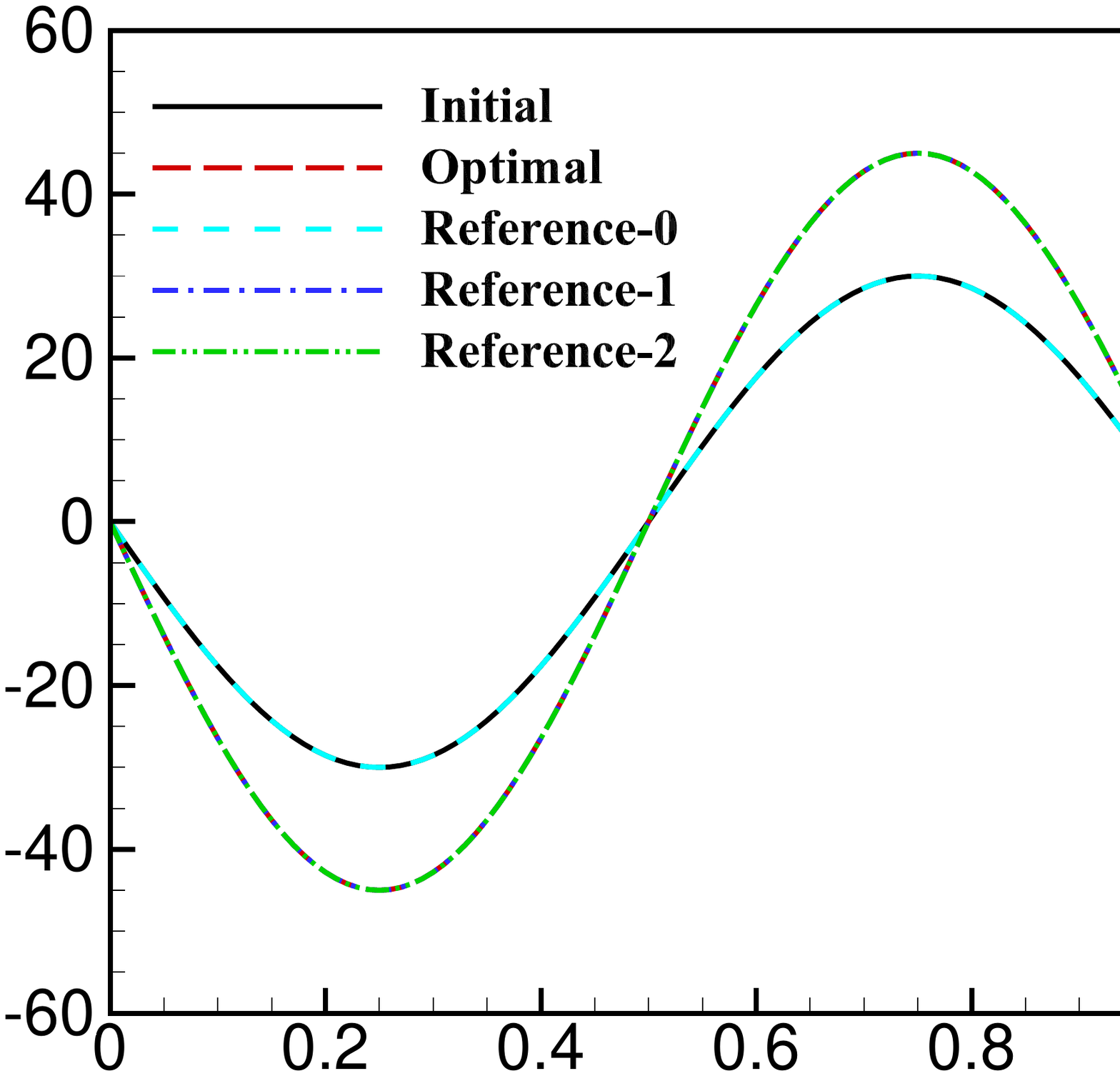}
\figlab 0.in -0.in {(b)}
\sethlabel {$t/T$}
\setvlabel {$\theta_{z^\prime}$ ($^\circ$)}
\includegraphics[width=0.35\textwidth]{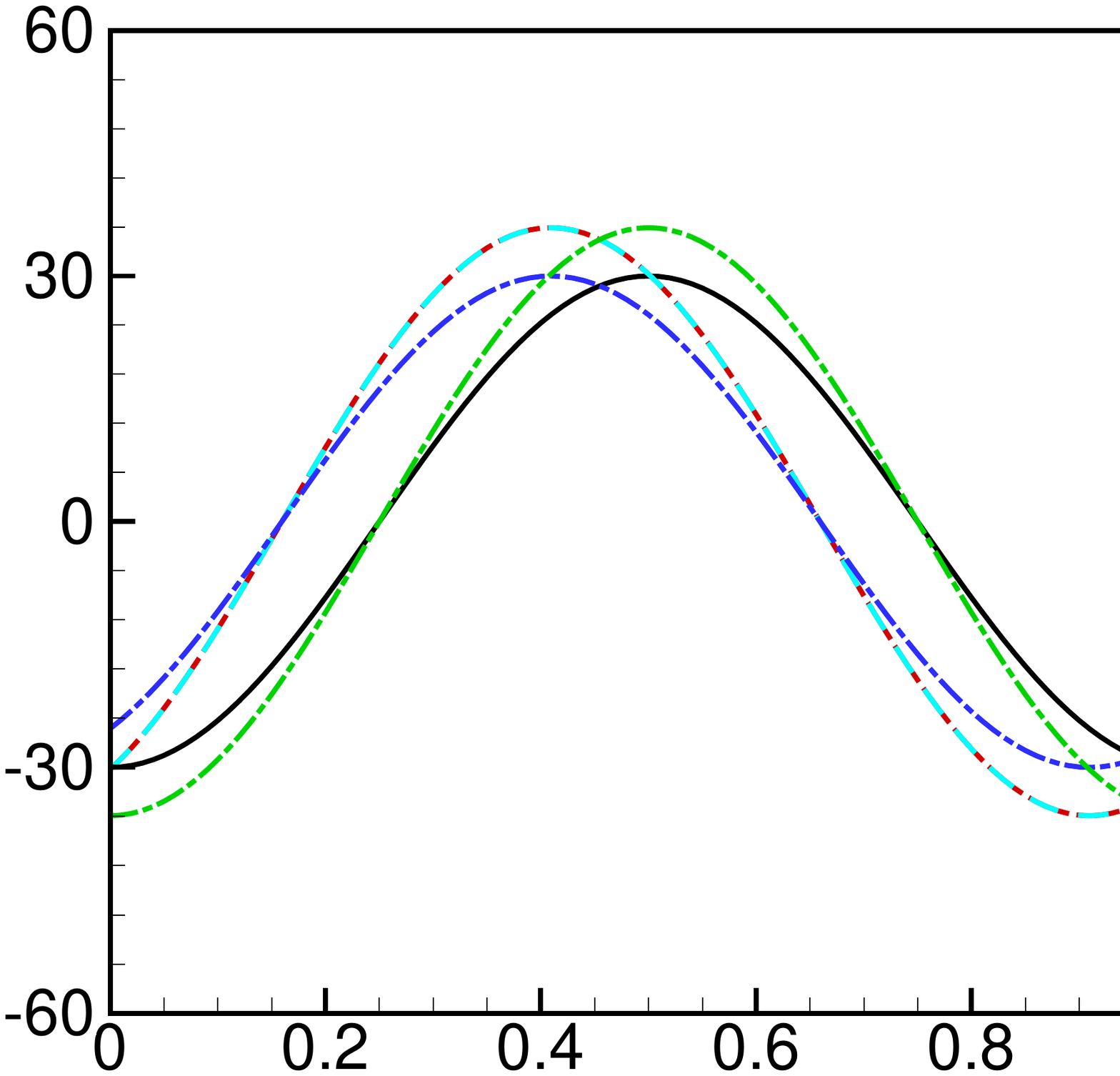}}
\centerline{
\figlab 0.in -0.in {(c)}
\sethlabel {$t/T$}
\setvlabel {$\alpha$ at 50\% span ($^\circ$)}
\includegraphics[width=0.35\textwidth]{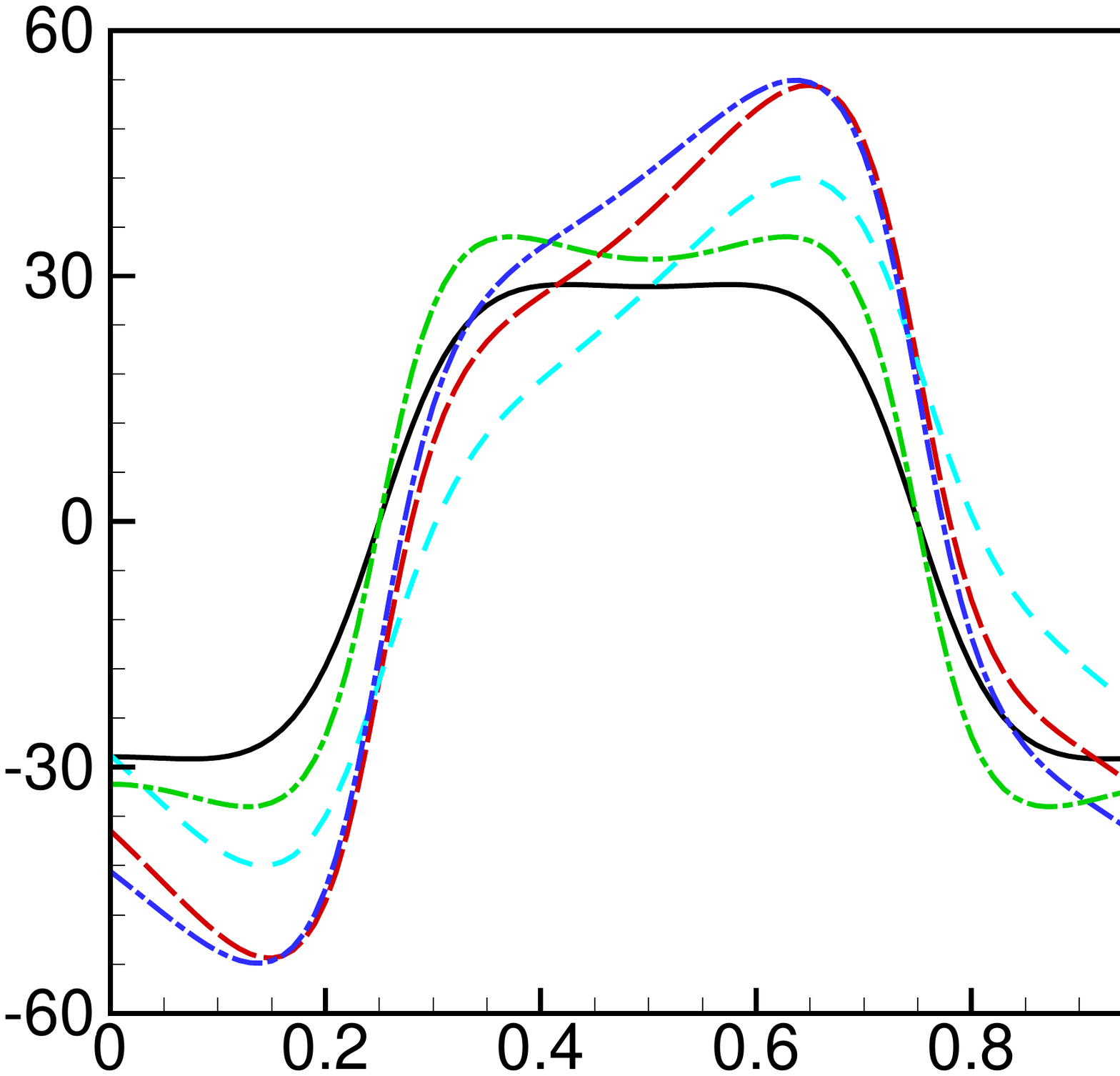}
\figlab 0.in -0.in {(d)}
\sethlabel {$t/T$}
\setvlabel {$\alpha$ at 70\% span ($^\circ$)}
\includegraphics[width=0.35\textwidth]{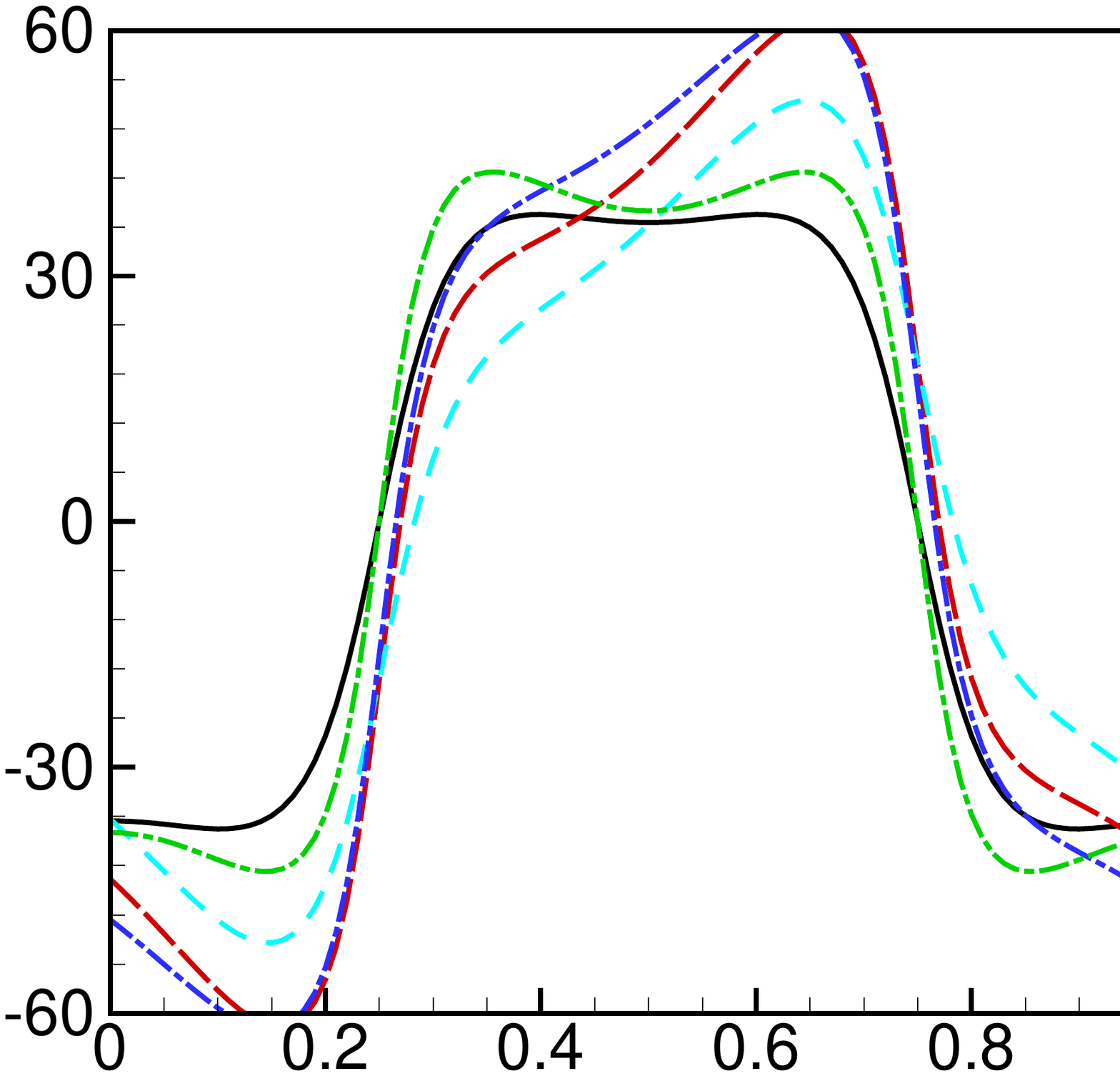}}
\centerline{
\figlab 0.in -0.in {(e)}
\includegraphics[width=0.7\textwidth]{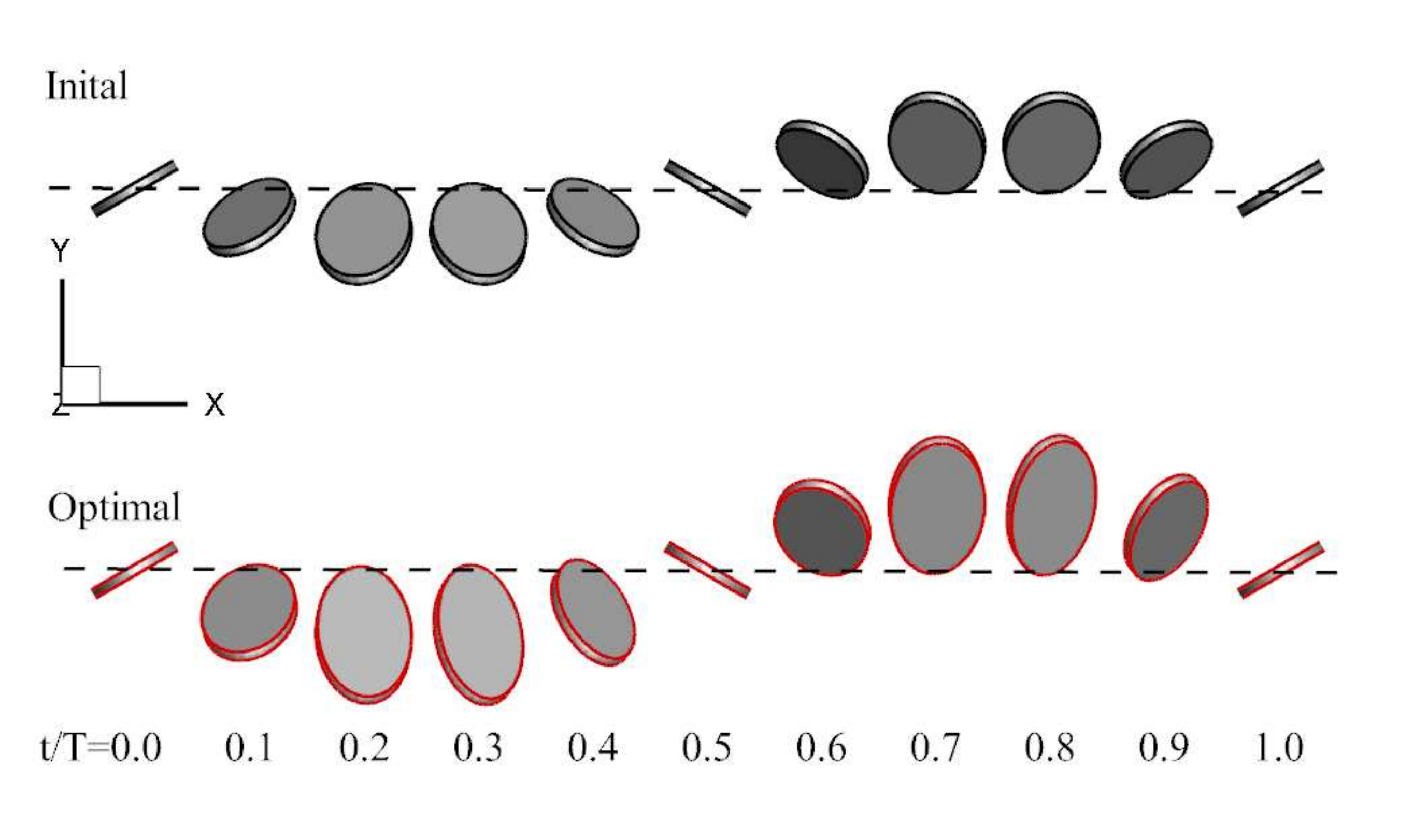}}
\caption{Comparison of the instantaneous rolling angle (a), pitching angle (b), angle of attack at 50\% span (c), and angle of attack at 70\% span (d), along with (e) the instantaneous kinematics comparison between the initial and optimal cases from the side view.}
\label{fig:angles}
\end{figure}

\begin{figure}
\centerline{
\sethlabel {$t/T$}
\setvlabel {${\bar C}_T$}
\includegraphics[width=0.4\textwidth]{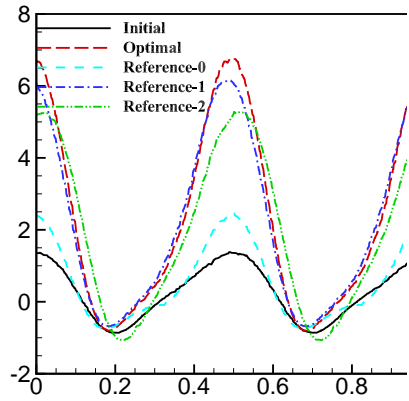}}
\caption{Comparison of the instantaneous thrust coefficients of the pitching-rolling plate with initial, optimal, reference-0, reference-1, and reference-2 controls}
\label{fig:3D_ForceComp}
\end{figure}

After 5 main iterations in the described adjoint-based optimization, the converged optimal control provides a dramatic boost of the thrust coefficient from 0.286 to 2.390 as it is shown in Table \ref{tab:drag_3D}. For the purpose of analysis, three reference cases in the table are simulated for comparison. Reference-0 uses the initial rolling amplitude and the optimal values for the other two parameters; Reference-1 uses the initial pitching amplitude and the optimal values for the other two; and Reference-2 uses the initial phase delay and the optimal values for the amplitudes.
The corresponding flapping kinematics and instantaneous thrust coefficients for the initial, the optimal, and the three reference cases are respectively plotted in Figure  \ref{fig:angles} and Figure  \ref{fig:3D_ForceComp}.

\begin{figure}
\leftline{
\figlab 0.in -0.1in {(a)}
\sethlabel {Initial}
\includegraphics[width=0.25\textwidth]{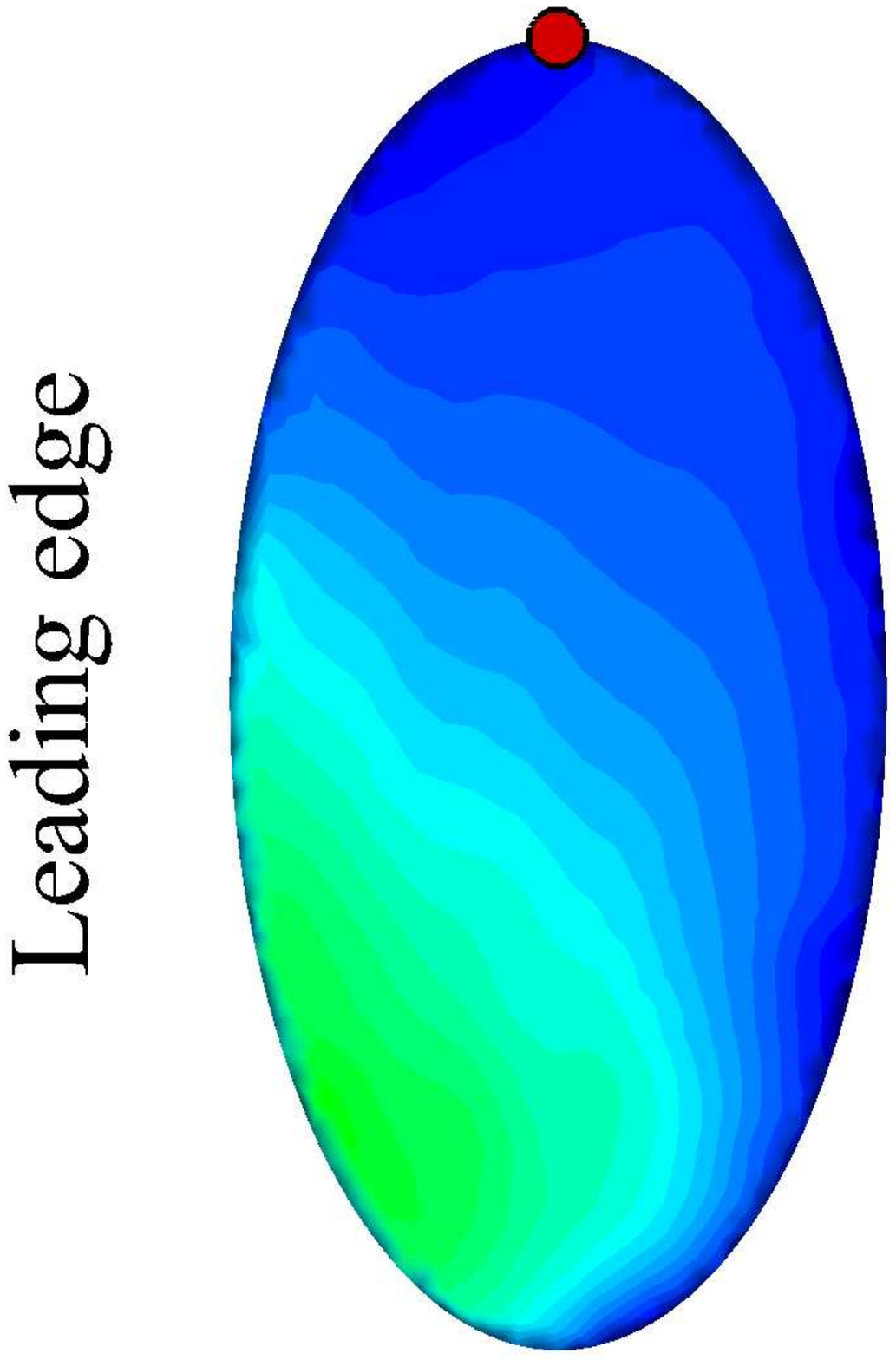}
\figlab 0.in -0.1in {(b)}
\sethlabel {Optimal}
\includegraphics[width=0.25\textwidth]{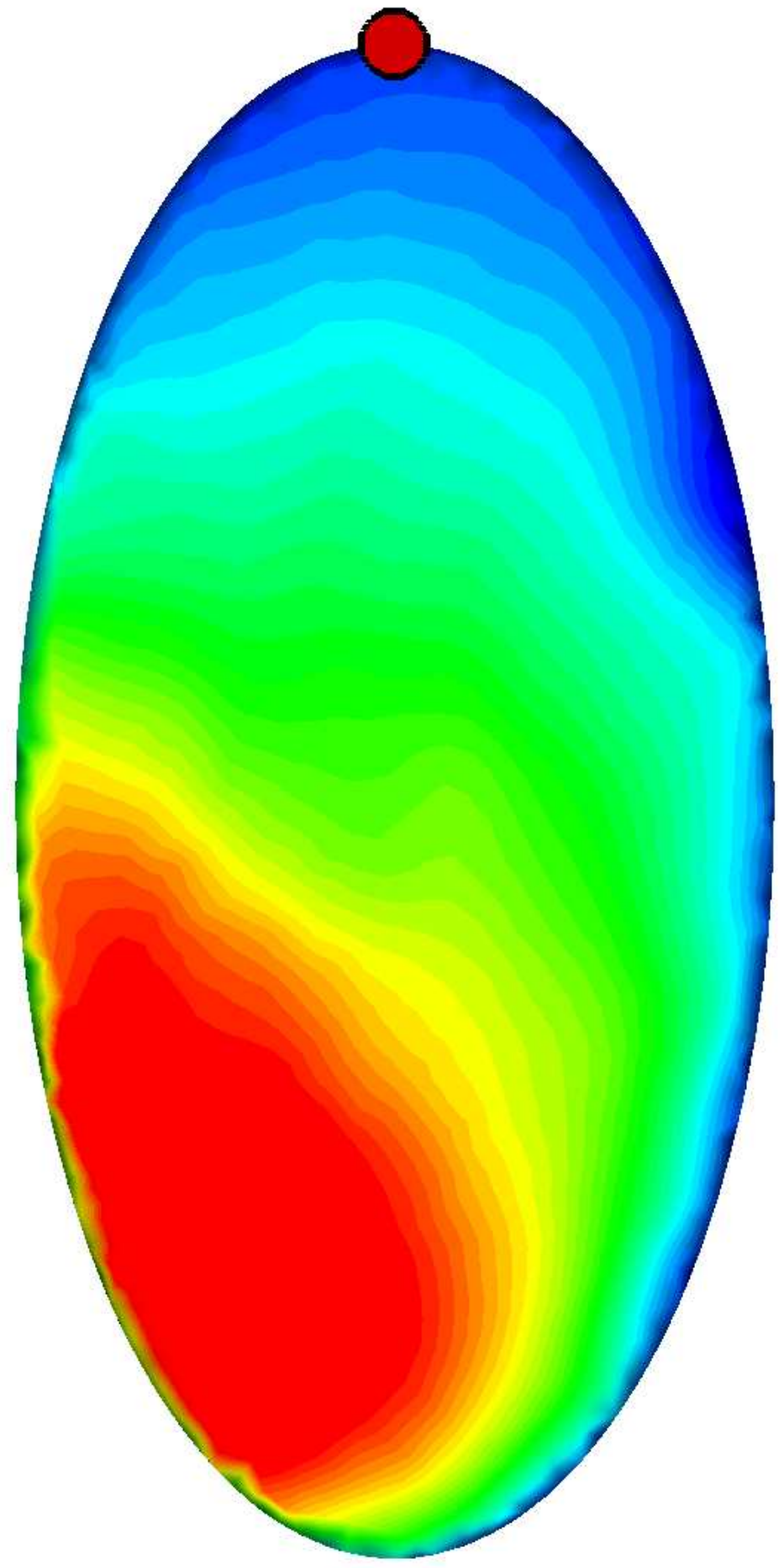}}
\leftline{
\sethlabel {Reference-0}
\figlab 0.in -0.1in {(c)}
\includegraphics[width=0.25\textwidth]{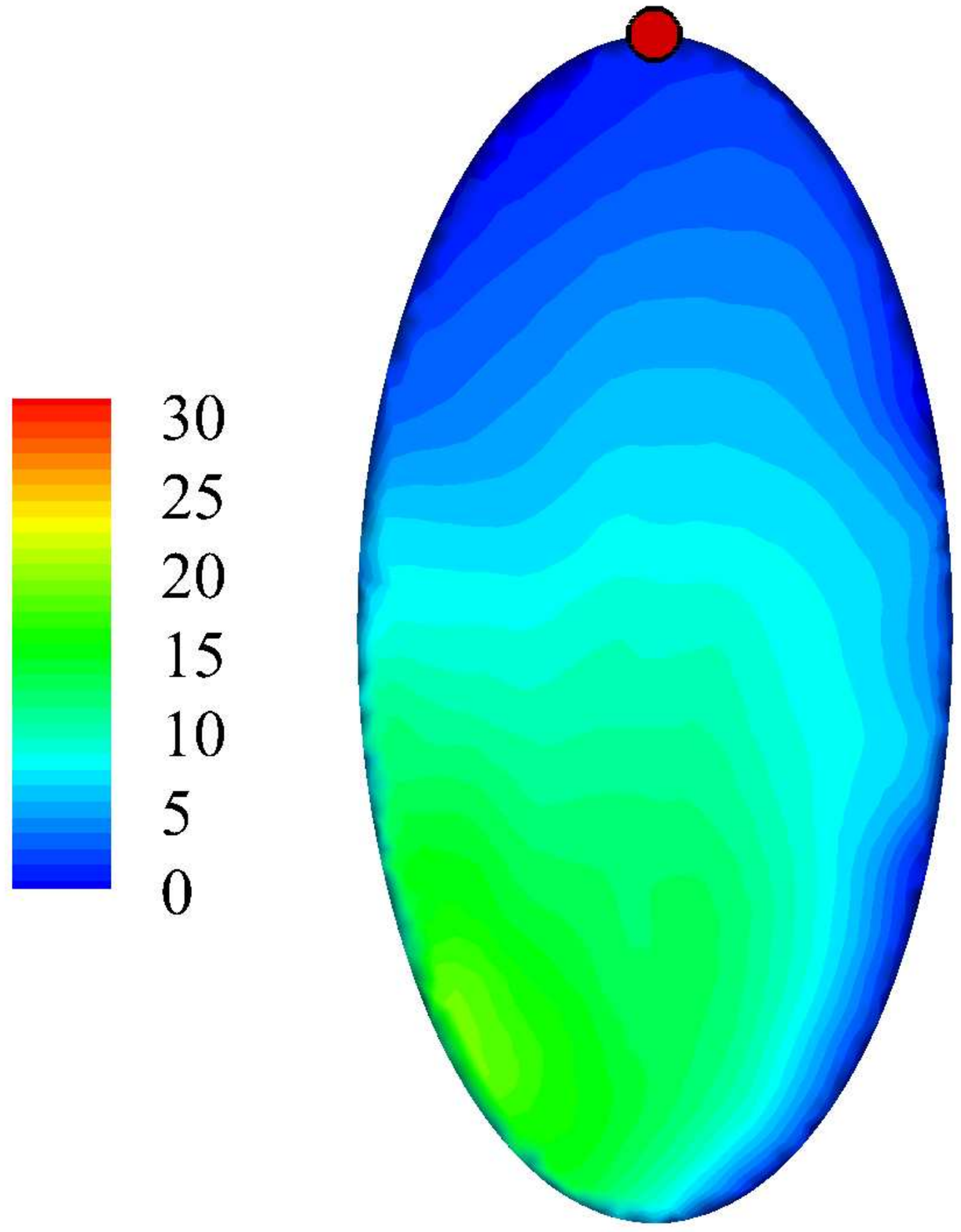}
\figlab 0.in -0.1in {(d)}
\sethlabel {Reference-1}
\includegraphics[width=0.25\textwidth]{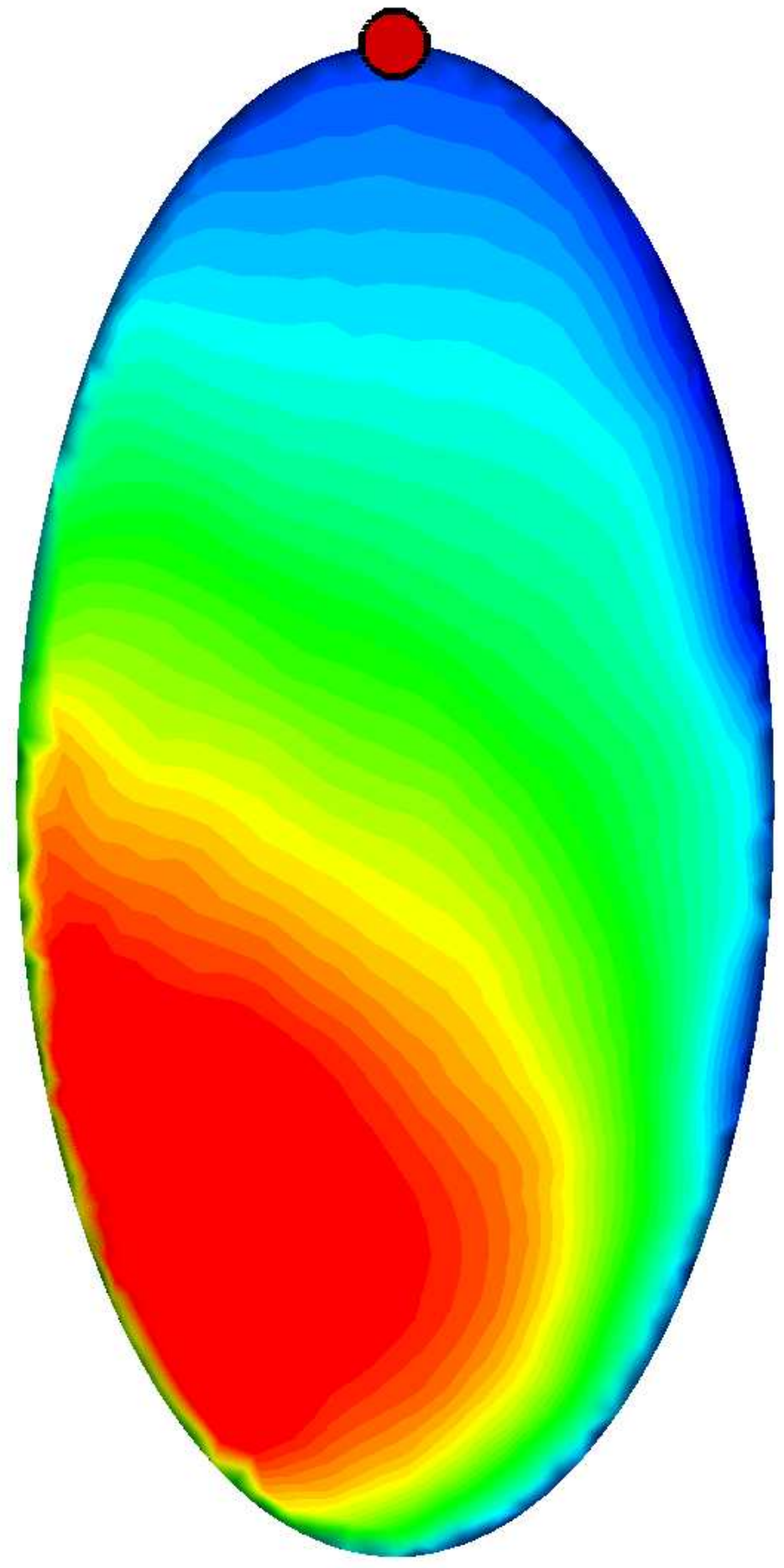}
\figlab 0.in -0.1in {(e)}
\sethlabel {Reference-2}
\includegraphics[width=0.25\textwidth]{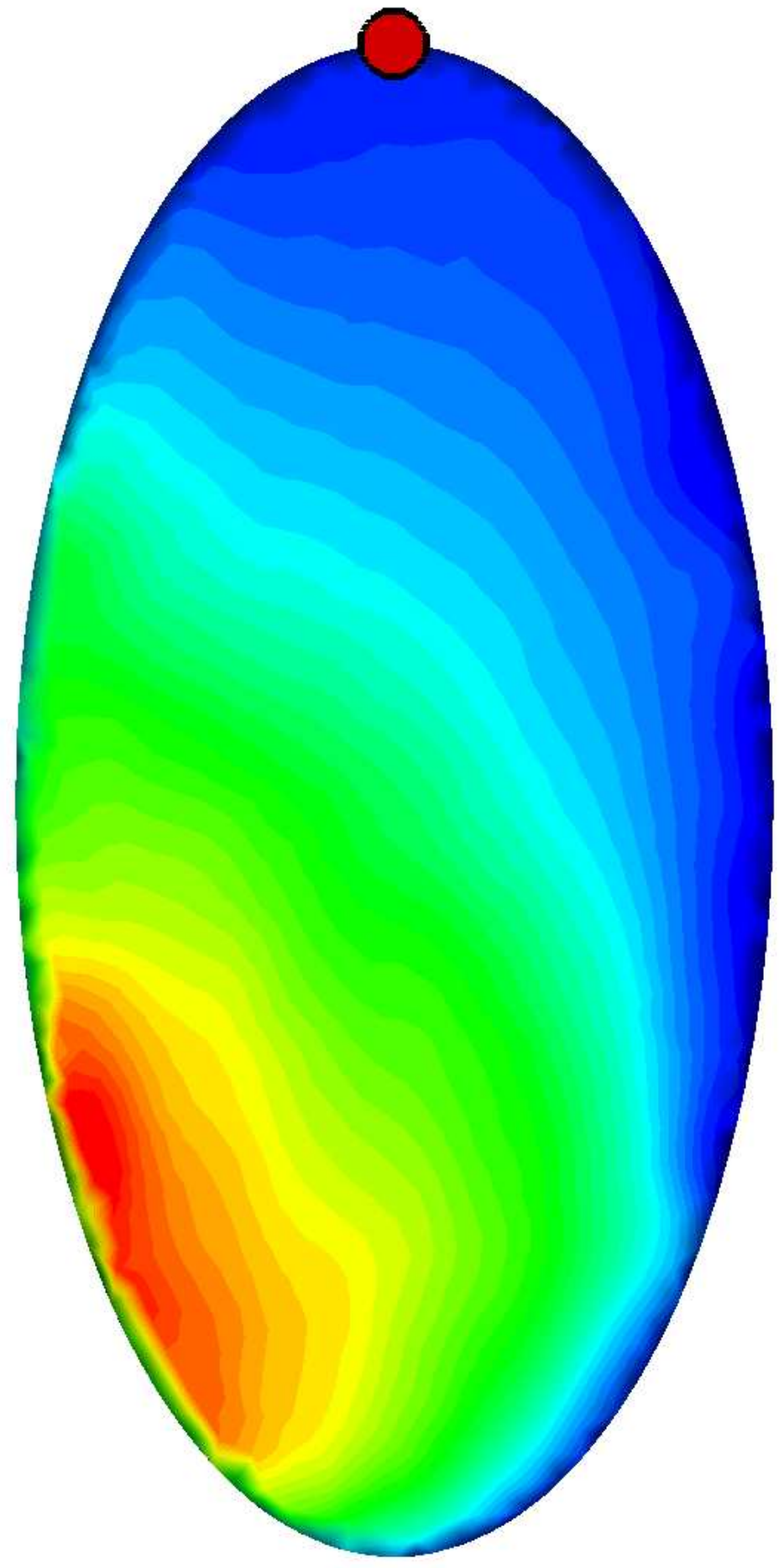}}
\caption{The pressure difference between the upper and lower surface of the pitching-rolling plate at t/T = 0.5 for the (a) initial, (b) optimal, (c) reference-0, (d) reference-1, and (e) reference-2 controls. The red dots indicate the location of wing root.}
\label{fig:pressure}
\end{figure}

The optimization increases the rolling amplitude to the upper bound limit. This contributes to the major improvement in the thrust optimization. If the initial rolling amplitude is used to replace the value in the optimal control, the thrust coefficient will be reduced significantly from 2.390 to 0.580, close to the value of 0.286 with the initial control. The increase in the rolling amplitude increases the rolling flapping velocity and the angle of attack. Between the two effects, the increase of flapping velocity increases both the plunging velocity and the circulation along the plate surface,  the increase of the angle of attack may enhance the circulation as well, and both contribute to the lift-induced thrust according to the analysis using quasi-steady model \cite{xu2016using,andersen2005unsteady}. The increase of rolling amplitude can also be interpreted as an increase in the Strouhal number. By amplifying the rolling amplitude from $30^\circ$ to $45^\circ$, the Strouhal number increases from 0.52 to 0.79. According to a recent study \cite{floryan2017scaling}, the total thrust, combing lift-based and added mass forces, can be modelled as a function of Strouhal number, which increases with the growth of Strouhal number. Our optimization result is consistent with this observation. The adjoint-based optimization also increases the pitching amplitude $a_z$ from $30^\circ$ to $35.9^\circ$, which only improves the propulsion slightly, as shown in Table \ref{tab:drag_3D}. The comparison between the optimal case and the reference-1 case show the change in pitching amplitude leads to a decrease in the magnitude of angle of attack  and an increase in the magnitude of pitching angle (Figure  \ref{fig:angles}). 
The phase delay between pitching and rolling is increased from $90^\circ$ to $122.6^\circ$ during the optimization. 
Table \ref{tab:drag_3D} shows that the thrust coefficient drops down from 2.390 to 1.991 by 16.7\%, if we change the phase delay in the optimal case back to the initial value. The phase delay changes the timing between pitching and rolling motion. It does not only changes the amplitude of the angle of attack, but also changes the timing when the maximum large angle occurs, as clearly shown in Figure  \ref{fig:angles}.

Figure  \ref{fig:3D_ForceComp} compares the time history of thrust coefficient for all the cases. The peak of thrust occurs at about $t/T = 0.5$, where the plate is at the middle rolling position and has the largest rolling velocity. Small drag is produced when the plate is at the lowest/highest rolling position and has smallest rolling velocity. Figure  \ref{fig:pressure} compares the pressure difference between the upper and the lower surfaces of all the cases at the thrust peak ($t/T = 0.5$). Comparing to the initial control, the pressure difference between upper and lower surface is much higher in the optimal control case, which mainly due to the increase in rolling amplitude as shown in Figure  \ref{fig:pressure}. The highest pressure regions are located at the outer board of the plate, where the angle of attack is large and a strong LEV forms. The comparison between the optimal and the reference-1 controls shows that although the reference-1 case has higher pressure difference between the upper and lower surfaces, the large pitching angle at $t/T = 0.5$ in the optimal case gives a larger projection area in the streamwise direction, which results in an overall larger thrust. A further increase of the pitching amplitude can induce a even larger projection area but it also reduces the angle of attack and LEV, causing less pressure difference. Therefore the optimal control provides a delicate balance between the two conflict factors. The reference-2 case owns the same pitching and rolling amplitudes with the optimal case, but has a much lower pressure difference compared to the optimal case. 
This indicates that the phase delay plays a another key role to modulate the pressure distribution on the propulsor's surface, aside from rolling amplitude.

\subsection{Comparison of vortex structures}

\begin{figure}
\centerline{
\includegraphics[width=0.85\textwidth]{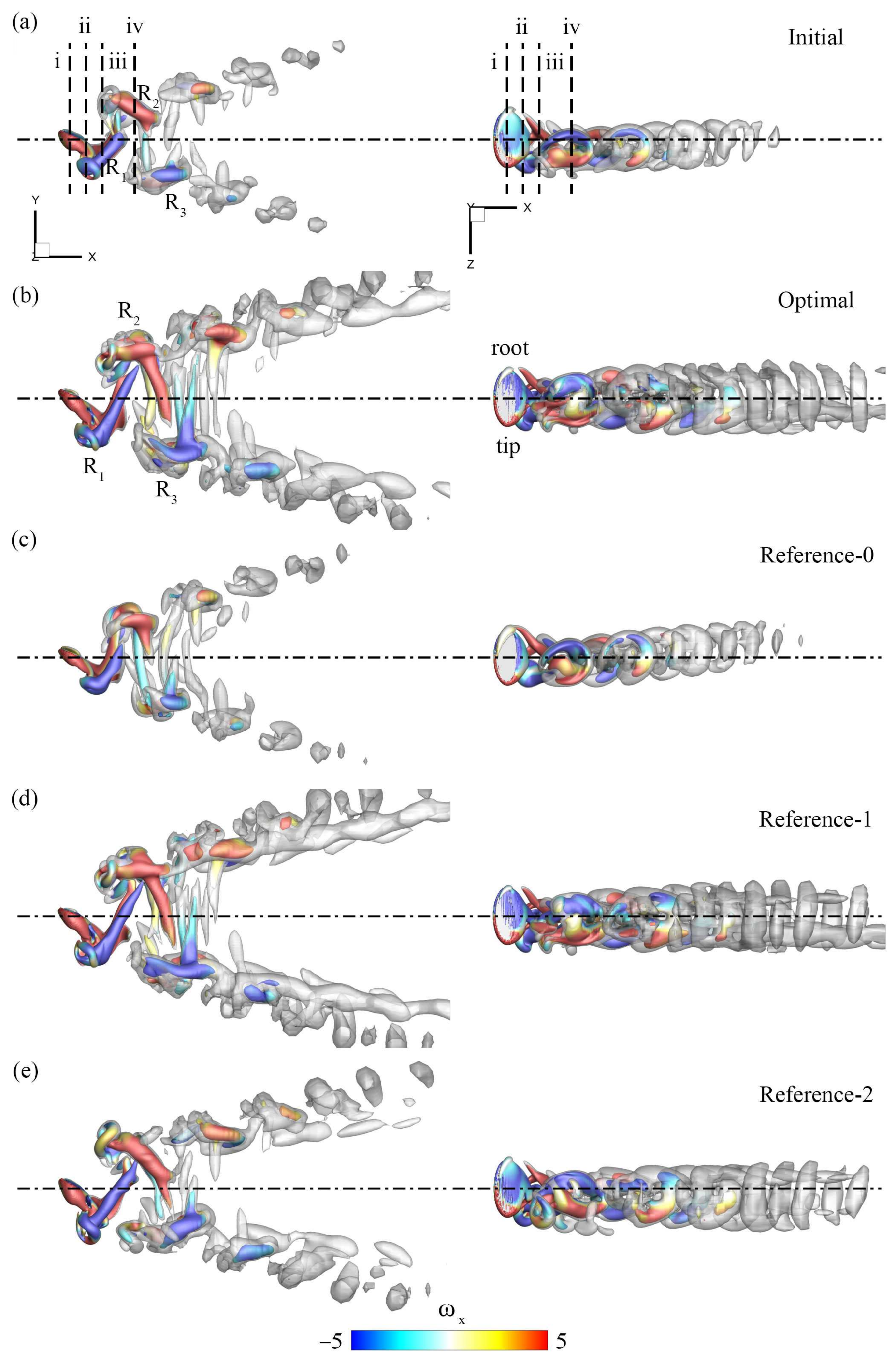}}
\caption{Comparison of the wake topology of the pitching-rolling plate with the (a) initial, (b) optimal, (c) reference-0, (d) reference-1, and (e) reference-2 controls at $t/T=0.5$.}
\label{fig:wake}
\end{figure}

\begin{figure}
\centerline{
\includegraphics[width=0.85\textwidth]{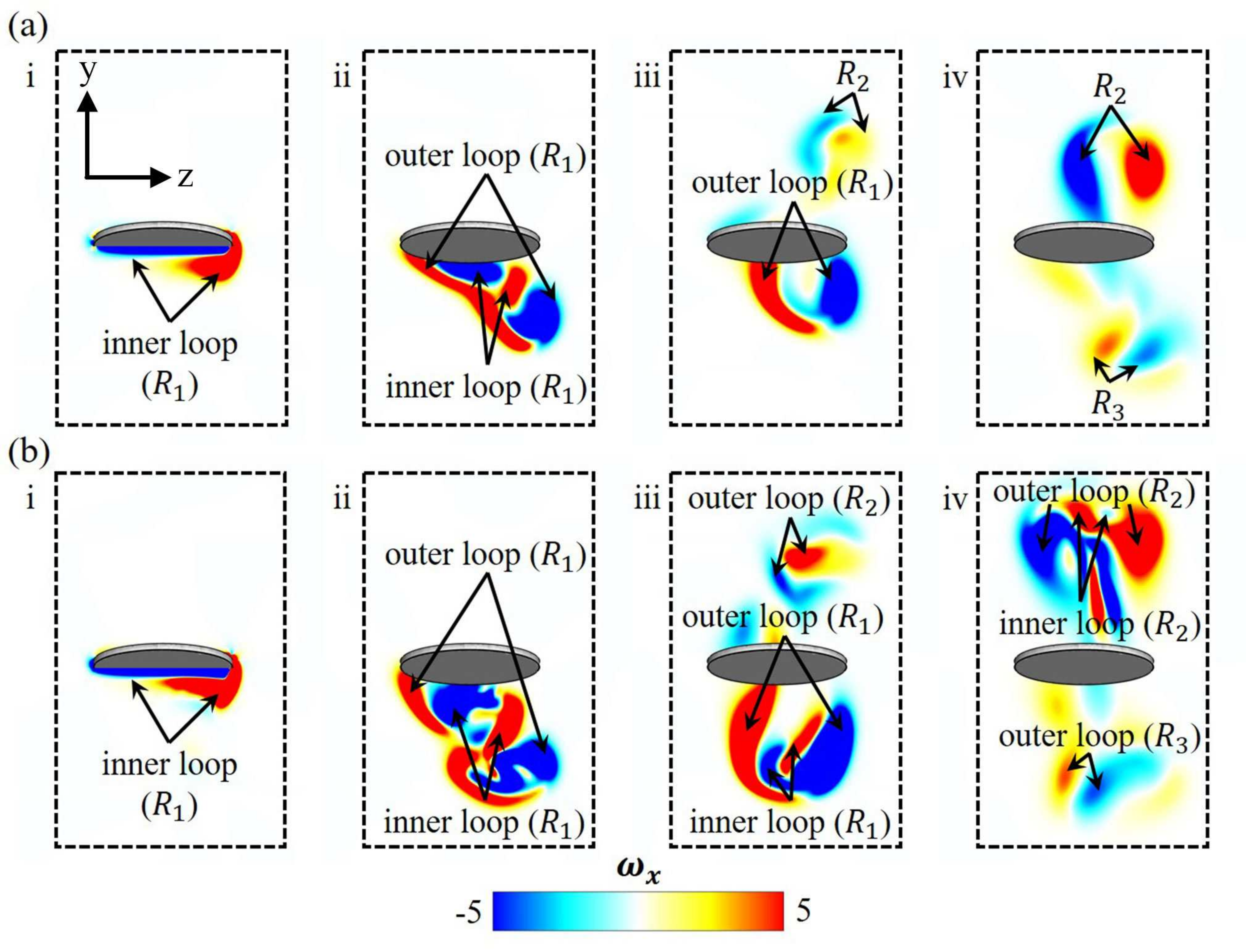}}
\caption{Slides contours of the streamwise vorticity $\omega_x$ for the (a) initial and (b) optimal controls at $t/T=0.5$. The spatial location of the slice cuts are marked as dashed lines (i-iv) in Figure \ref{fig:wake} (a) as (i) $x=0$, (ii) $x=0.6c$, (iii) $x=1.2c$, and (iv) $x=2.4c$.}
\label{fig:streamwise-vorticity}
\end{figure}

Figure  \ref{fig:wake} compares the wake topology of each cases. During the pitching-rolling motion, a pair of vortex rings are produced from each flapping cycle and form a bifurcated wake pattern in the downstream. For the initial control (Figure ~\ref{fig:wake}a), the downstream vortex rings gradually become weaker and annihilate quickly due to the viscous dissipation effect at such a low Reynolds number. The optimal control (Figure  \ref{fig:wake}b), however, enhances the strength of the vortex rings, and makes the shed vortex rings propagate a longer distance in the downstream without annihilation. Another noticeable difference in the wake topology is  wake deflection: the vortex streets in the high thrust cases have much smaller wake deflection in downstream, which make the flow prolusion all contribute to thrust rather than the lateral force.


Figure  \ref{fig:streamwise-vorticity} compares the 2D streamwise vorticity contours between the initial and optimal controls in the near wake. The spatial location of the slice cuts are labeled in Figure  \ref{fig:wake} (a). Red and blue indicate the counterclockwise and clockwise rotation direction, respectively. In Ref \cite{li2016three}, Li and Dong found that the vortex structures generated by a pitching-rolling plate contain both inner and outer vortex loops, which is different from the single loop vortex structures observed in pitching-heaving foils. This unique vortex formation is named double-C shaped vortex in their study. In the present work, similar vortex structures are also observed. Figure  \ref{fig:streamwise-vorticity} (i-ii) show the structures of the inner and outer loops of $R_1$. For the initial control, the double-C shaped vortex structure rapidly evolves into a single-loop vortex ring ($R_2$ , Figure  \ref{fig:streamwise-vorticity} a (iii)) as it convects downstream (Figure  \ref{fig:streamwise-vorticity} a (iii-iv)). For the optimal control, however, the enhancement of vortex strength makes double-C shaped vortex structure last for a longer period in the downstream. Both inner and outer loops of $R_2$ are still visible for the optimal case (Figure  \ref{fig:streamwise-vorticity} b (iii-iv)).

\begin{figure}
\centerline{
\includegraphics[width=0.85\textwidth]{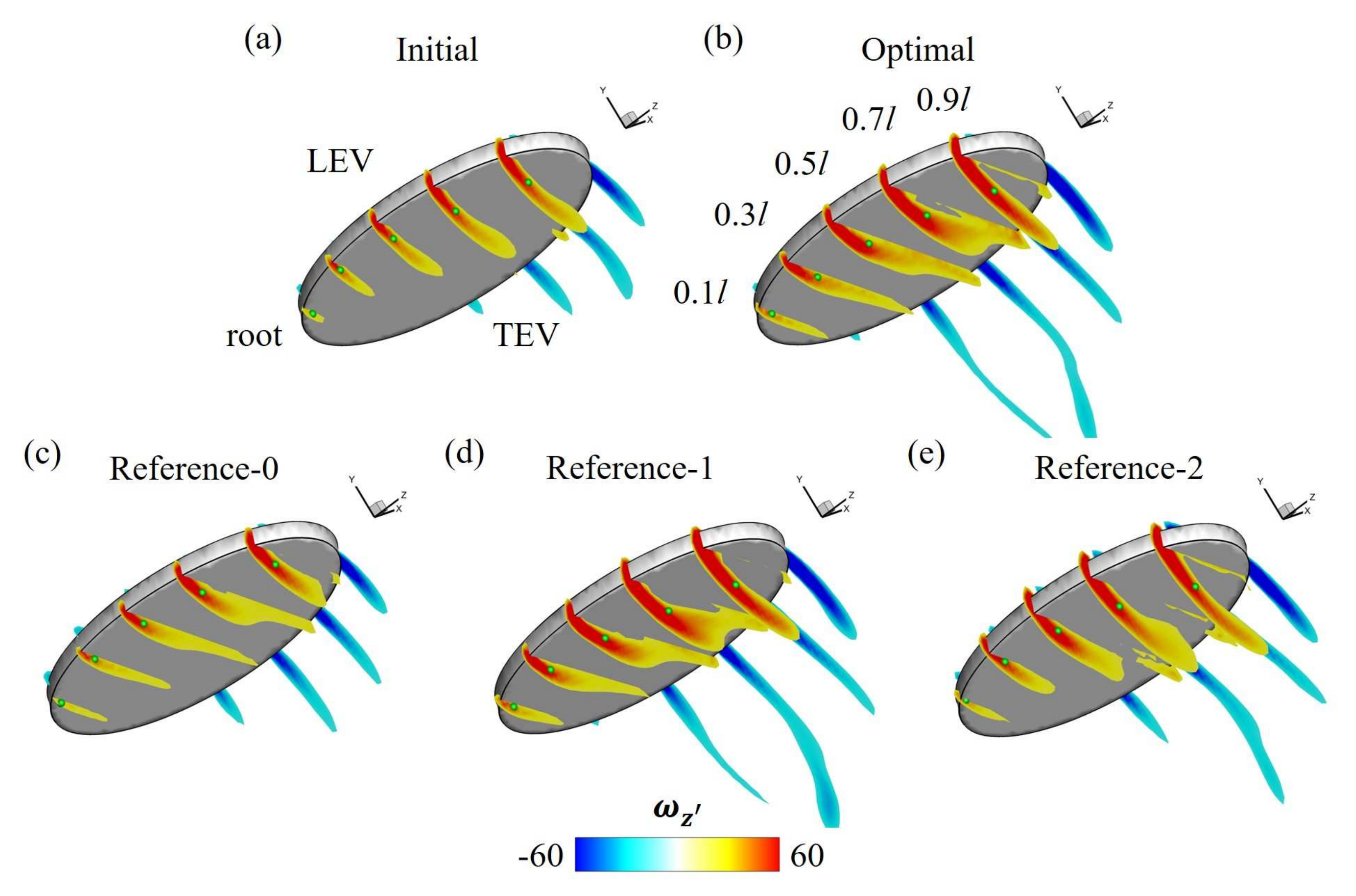}}
\caption{Comparison of the spanwise vorticity contour of the pitching-rolling plate with the (a) initial, (b) optimal, (c) reference-0, (d) reference-1, and (e) reference-2 controls at $t/T=0.5$. The slices are taken from the wing root to wing tip. The corresponding vortex center are marked with green dots at each slice.}
\label{fig:spanwise-vorticity}
\end{figure}

\begin{figure}
\centerline{
\includegraphics[width=0.5\textwidth]{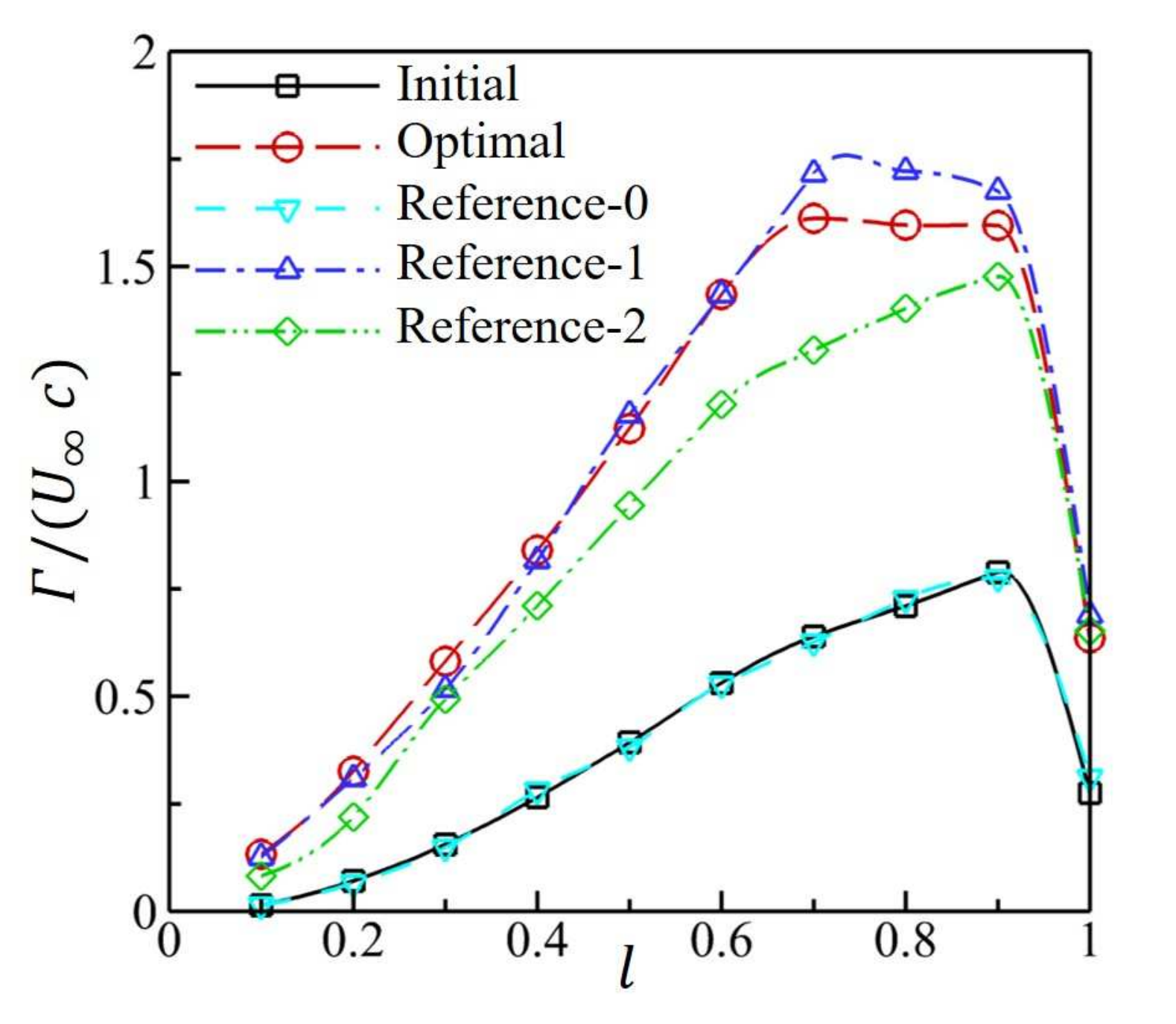}}
\caption{Comparison of LEV circulation along the wing span at $t/T = 0.5$. The vortex circulation is normalized by $U_\infty c$.}
\label{fig:circulation}
\end{figure}

\begin{figure}
\centerline{
\includegraphics[width=0.5\textwidth]{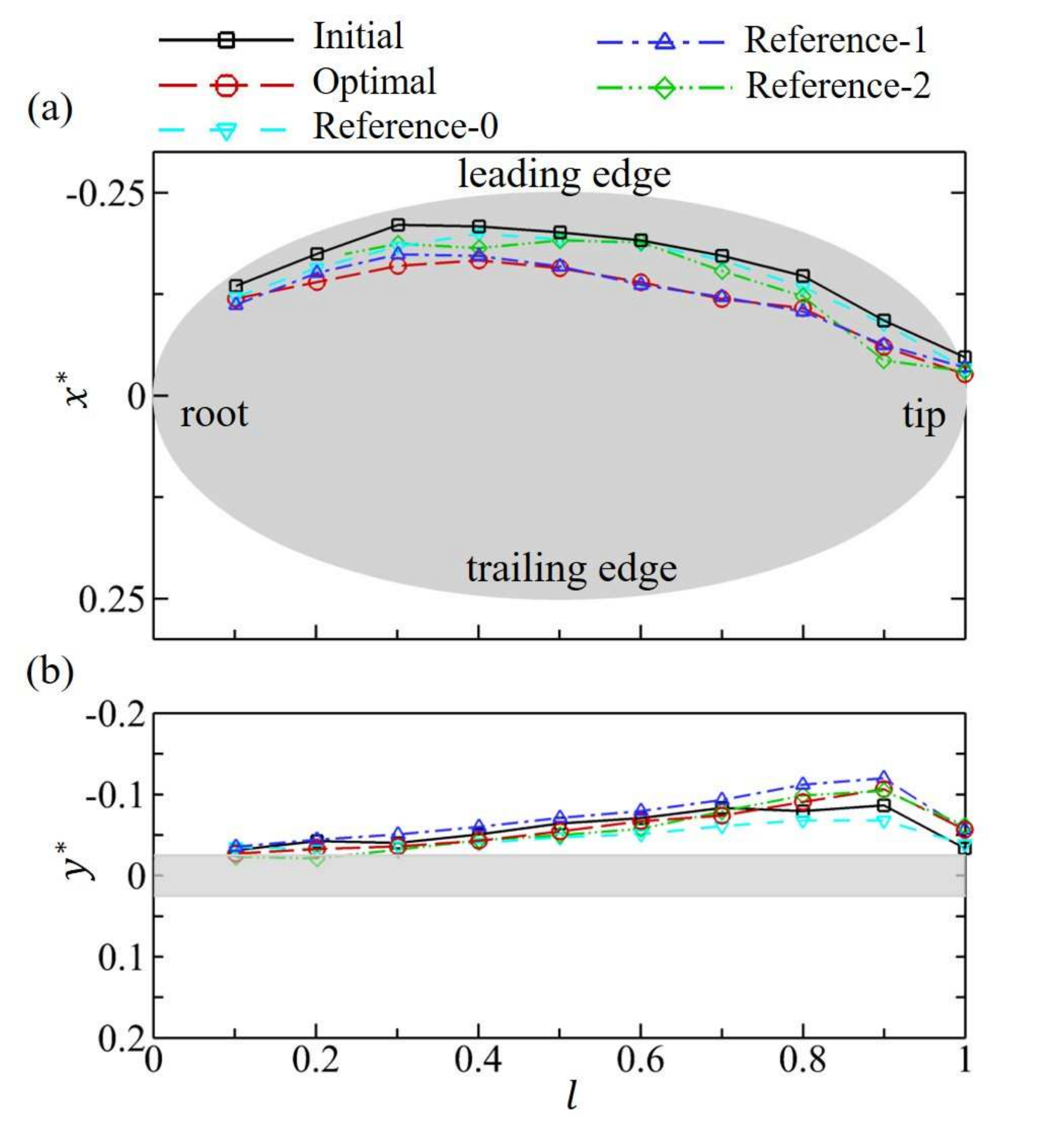}}
\caption{Comparison of LEV position along the wing span at $t/T = 0.5$: (a) the position from leading edge $x^\ast$, (b) the position above wing surface $y^\ast$.}
\label{fig:LEVPositions}
\end{figure}

Next, the LEV and trailing-edge vortex (TEV) formations are compared for the cases with initial, optimal, reference-0, reference-1 and reference-2 controls at $t/T = 0.5$. For each case, multiple 2D slide cuts are taken along the wingspan. Five slices are shown in Figure  \ref{fig:spanwise-vorticity} from 10\% to 90\% along the spanwise direction from the wing root to the wing tip. The size of the LEV continues to grow proportionally along the span. Quantitative measurement of LEV circulation distributions along the wingspan are performed based on the 2D flow slices. The circulation is calculated based on the spanwise vorticity ($\omega_{z^\prime}$ ) contours, and then normalized by $U_\infty c$: we first identify a closed contour line around the vortex with a specified level ($\omega_{z^\prime}=34$), and the circulation ($\Gamma$) is then calculated by integrating along this line. Although the magnitude of the circulation depend on the chosen contour level, the characteristic behavior of the vortex is not affected by this choice. Figure  \ref{fig:circulation} compares the normalized circulation of the LEVs along the wingspan. Both optimal and reference controls generate much stronger LEVs compared to that of the initial control due to the larger rolling amplitude. As a result of the phase delay angle adjustment, the LEV circulations of optimal and reference-1 controls are further enhanced compared to that of the reference-2 control. The amplitudes of LEV circulations at difference control cases are consistent with the angle of attacks shown in Figure  \ref{fig:angles}.


The LEV center of each slide is also determined based on the vortex contour shape in the same way as the LEV circulation. The distance between the vortex center (green dots in Figure  \ref{fig:spanwise-vorticity}) and plate surface, which are typically named as LEV lift-off distance, are measured and visualized. Figure  \ref{fig:LEVPositions} presents the chordwise LEV center position and the lift-off height above the plate surface for each control case at $t/T = 0.5$. 
The lift-off height increases along with the strength of LEV. Too large LEV may result in the detachment of LEV, thus a reduction in the thrust performance. Therefore, there should be a balance between enhancing the LEV strength and maintaining the LEV attachment for thrust improvement.

\section{Conclusion} \label{sec:conclusion}
The propulsion performance of a pitching-rolling plate has been investigated using an adjoint-based optimization approach. The rolling amplitude, the pitching amplitude, and the phase delay between the pitching and rolling motion are chosen as  control parameters to be optimized for thrust performance. After five main design iterations, the thrust coefficient increases from 0.286 to 2.390 by approximately eight times larger. The improvement mainly results from the increase of the rolling amplitude and the adjustment of the phase delay. The higher rolling amplitude enhances the pressure difference between the upper and lower surfaces, and thus improves the thrust generation. It contributes to the major improvement in the thrust optimization. On the other hand, the adjustment of the phase delay improves the thrust coefficient by 16.7\%. 
The wake structure analysis shows that the optimal control enhances the strength of the vortex street, and reduce the wake defect. Most thrust is generated around the plate’s tip and leading-edge regions due to the formation of the LEV. 

\section*{Appendix} \label{sec:appendix}
Assuming that the linearized equation is governed by linearized Navier-Stokes equation with infinitesimal body force or infinitesimal mass source $\mathbf{f}$ as the source term, and the control is the infinitesimal body force or mass source instead of solid motion, then the linearized equation can be formulated as
\begin{equation}\label{eqn:linear_eqn1}
\begin{aligned}
\mathcal N^\prime(\mathbf q){\mathbf q^\prime} &= \mathbf{f^\prime} \qquad \mathrm{in}\;\; \Omega,\\
{\mathbf u^\prime} &= 0 \qquad \mathrm{on} \;\; \mathcal S, \; \Gamma_\infty, \\
\frac{\partial p^\prime} {\partial n}  &= 0 \qquad  \mathrm{on} \;\; \Gamma_\infty.\\
\end{aligned}
\end{equation}
The derivative of the cost function is
\begin{equation}
\begin{aligned}
\mathcal J^\prime =  -\frac {1}{T D_0} \left (\int_0^T \int_{\mathcal S} {\boldsymbol \sigma_1}^\prime \cdot {\mathbf n}  \text{d}s \text{d}t + \int_0^T \int_{\Omega} {\mathbf q}^\ast \cdot [ \mathcal N^\prime(\mathbf q) \mathbf q^\prime - {\mathbf f}^\prime] \text{d}\Omega \text{d}t \right).
\end{aligned}
\end{equation}
Following the same derivation as in section \ref{sec:adjoint_equation}, we get
\begin{equation}
\begin{aligned}
\mathcal J^\prime =  -\frac {1}{T D_0}\left( b -  \int_0^T \int_{\Omega} {\mathbf q}^\prime \cdot \mathcal N^\ast(\mathbf q) \mathbf q^\ast \text{d}\Omega \text{d}t - \int_0^T \int_{\Omega} {\mathbf q}^\ast  \cdot {\mathbf f}^\prime \text{d}\Omega \text{d}t\right),
\end{aligned}
\end{equation}
with $\mathcal N^\ast(\mathbf q) \mathbf q^\ast$ being the same as that in equation (\ref{eqn:NS-adj}) and
\begin{equation}
\begin{aligned}
b =   \left. \int_{\Omega} u^\ast_j u^\prime_j \text{d}\Omega \right|_{t=0}^{t=T} + b_{\infty} + \int_0^T \int_{\Omega} (u^\ast_i+\delta_{1i})\sigma^\prime_{ij}n_j \text{d}s\text{d}t - \int_0^T \int_{\Omega} u_i^\prime(\sigma_{ij}^\ast n_j+u_j^\ast u_j n_i) \text{d}s\text{d}t.
\end{aligned}
\end{equation}
When the adjoint equation (\ref{eqn:ad_eqn1}) is satisfied, the derivative of the cost function reduces to
\begin{equation}
\begin{aligned}
\mathcal J^\prime =  \frac {1}{T D_0}\left( \int_0^T \int_{\Omega} {\mathbf q}^\ast  \cdot {\mathbf f}^\prime \text{d}t\right).
\end{aligned}
\end{equation}
This reveals the physical meaning of the adjoint velocity and pressure, ${\mathbf q}^\ast$. They are the transfer functions from body force or mass source perturbation, ${\mathbf f}^\prime$, to the perturbed cost function, $\mathcal J^\prime$. These fields show how a change in the body force and mass source can directly affect the cost function.

\section*{Acknowledgements} \label{sec:acknowledgements}
This research has been supported by U.S. Army Research Laboratory (ARL) through MAST CTA W911NF-08-2-0004, AFOSR grant FA9550-12-1-0071, NSF grant CBET-1313217, and ONR MURI grant N00014-14-1-0533.


\begin{thebibliography}{40}
\newcommand{\enquote}[1]{``#1''}
\providecommand{\natexlab}[1]{#1}
\providecommand{\url}[1]{\texttt{#1}}
\providecommand{\urlprefix}{URL }
\expandafter\ifx\csname urlstyle\endcsname\relax
  \providecommand{\doi}[1]{doi:\discretionary{}{}{}#1}\else
  \providecommand{\doi}{doi:\discretionary{}{}{}\begingroup
  \urlstyle{rm}\Url}\fi

\bibitem[{Mueller and DeLaurier(2001)}]{mueller2001overview}
Mueller, T.~J., and DeLaurier, J.~D., \enquote{An overview of micro air vehicle
  aerodynamics,} \emph{Fixed and flapping wing aerodynamics for micro air
  vehicle applications}, Vol. 195, 2001, pp. 1--9.
\newblock \doi{10.2514/5.9781600866654.0001.0010},
  \urlprefix\url{https://doi.org/10.2514%2F5.9781600866654.0001.0010}.

\bibitem[{Anderson et~al.(1998)Anderson, Streitlien, Barrett, and
  Triantafyllou}]{anderson1998oscillating}
Anderson, J., Streitlien, K., Barrett, D., and Triantafyllou, M.,
  \enquote{Oscillating foils of high propulsive efficiency,} \emph{J. Fluid
  Mech.}, Vol. 360, 1998, pp. 41--72.
\newblock \doi{10.1017/s0022112097008392},
  \urlprefix\url{https://doi.org/10.1017%2Fs0022112097008392}.

\bibitem[{Koochesfahani(1989)}]{koochesfahani1989vortical}
Koochesfahani, M.~M., \enquote{Vortical Patterns in the Wake of an Oscillating
  Airfoil,} \emph{AIAA J.}, Vol.~27, No.~9, 1989.
\newblock \doi{10.2514/3.10246},
  \urlprefix\url{https://doi.org/10.2514%2F3.10246}.

\bibitem[{Dong et~al.(2006)Dong, Mittal, and Najjar}]{Dong:2006}
Dong, H., Mittal, R., and Najjar, F., \enquote{Wake topology and hydrodynamic
  performance of low aspect-ratio flapping foils,} \emph{J. Fluid Mech.}, Vol.
  566, 2006, pp. 309--343.
\newblock \doi{10.1017/s002211200600190x},
  \urlprefix\url{https://doi.org/10.1017%2Fs002211200600190x}.

\bibitem[{Yang et~al.(2010)Yang, Wei, and Zhao}]{YangWeiZhao:2010}
Yang, T., Wei, M., and Zhao, H., \enquote{Numerical study of flexible flapping
  wing propulsion,} \emph{AIAA J.}, Vol.~48, No.~12, 2010, pp. 2909--2915.
\newblock \doi{10.2514/1.j050492},
  \urlprefix\url{https://doi.org/10.2514%2F1.j050492}.

\bibitem[{Bandyopadhyay et~al.(2012)Bandyopadhyay, Beal, Hrubes, and
  Mangalam}]{bandyopadhyay2012relationship}
Bandyopadhyay, P.~R., Beal, D.~N., Hrubes, J.~D., and Mangalam, A.,
  \enquote{Relationship of roll and pitch oscillations in a fin flapping at
  transitional to high Reynolds numbers,} \emph{J. Fluid Mech.}, Vol. 702,
  2012, pp. 298--331.
\newblock \doi{10.1017/jfm.2012.178},
  \urlprefix\url{https://doi.org/10.1017%2Fjfm.2012.178}.

\bibitem[{Techet(2008)}]{techet2008propulsive}
Techet, A.~H., \enquote{Propulsive performance of biologically inspired
  flapping foils at high Reynolds numbers,} \emph{J. Exp. Biol.}, Vol. 211,
  No.~2, 2008, pp. 274--279.
\newblock \doi{10.1016/s0021-9290(06)84420-9},
  \urlprefix\url{https://doi.org/10.1016%2Fs0021-9290%2806%2984420-9}.

\bibitem[{Li and Dong(2016)}]{li2016three}
Li, C., and Dong, H., \enquote{Three-dimensional wake topology and propulsive
  performance of low-aspect-ratio pitching-rolling plates,} \emph{Phys.
  Fluids}, Vol.~28, No.~7, 2016, p. 071901.
\newblock \doi{10.1063/1.4954505},
  \urlprefix\url{https://doi.org/10.1063%2F1.4954505}.

\bibitem[{Berman and Wang(2007)}]{berman2007energy}
Berman, G.~J., and Wang, Z.~J., \enquote{Energy-minimizing kinematics in
  hovering insect flight,} \emph{J. Fluid Mech.}, Vol. 582, No.~1, 2007, pp.
  153--168.
\newblock \doi{10.1017/s0022112007006209},
  \urlprefix\url{https://doi.org/10.1017%2Fs0022112007006209}.

\bibitem[{Ghommem et~al.(2012)Ghommem, Hajj, Mook, Stanford, Beran, Snyder, and
  Watson}]{ghommem2012global}
Ghommem, M., Hajj, M.~R., Mook, D.~T., Stanford, B.~K., Beran, P.~S., Snyder,
  R.~D., and Watson, L.~T., \enquote{Global optimization of actively morphing
  flapping wings,} \emph{Journal of Fluids and Structures}, Vol.~33, 2012, pp.
  210--228.
\newblock \doi{10.1016/j.jfluidstructs.2012.04.013},
  \urlprefix\url{https://doi.org/10.1016%2Fj.jfluidstructs.2012.04.013}.

\bibitem[{Milano and Gharib(2005)}]{milano2005uncovering}
Milano, M., and Gharib, M., \enquote{Uncovering the physics of flapping flat
  plates with artificial evolution,} \emph{J. Fluid Mech.}, Vol. 534, No.~1,
  2005, pp. 403--409.
\newblock \doi{10.1017/s0022112005004842},
  \urlprefix\url{https://doi.org/10.1017%2Fs0022112005004842}.

\bibitem[{Trizila et~al.(2011)Trizila, Kang, Aono, Shyy, and
  Visbal}]{trizila2011low}
Trizila, P., Kang, C.-K., Aono, H., Shyy, W., and Visbal, M.,
  \enquote{Low-Reynolds-number aerodynamics of a flapping rigid flat plate,}
  \emph{AIAA J.}, Vol.~49, No.~4, 2011, pp. 806--823.
\newblock \doi{10.2514/1.j050827},
  \urlprefix\url{https://doi.org/10.2514%2F1.j050827}.

\bibitem[{Bewley et~al.(2001)Bewley, Moin, and Temam}]{Bewley:2001}
Bewley, T.~R., Moin, P., and Temam, R., \enquote{{\uppercase{DNS}}-based
  predictive control of turbulence: an optimal benchmark for feedback
  algorithms,} \emph{J. Fluid Mech.}, Vol. 447, 2001, pp. 179--225.
\newblock \doi{10.1017/s0022112001005821},
  \urlprefix\url{https://doi.org/10.1017%2Fs0022112001005821}.

\bibitem[{Wei and Freund(2006)}]{Wei:2006}
Wei, M., and Freund, J.~B., \enquote{A noise-controlled free shear flow,}
  \emph{J. Fluid Mech.}, Vol. 546, 2006, pp. 123--152.
\newblock \doi{10.1017/s0022112005007093},
  \urlprefix\url{https://doi.org/10.1017%2Fs0022112005007093}.

\bibitem[{Jones and Yamaleev(2013)}]{jones2013adjoint}
Jones, M., and Yamaleev, N.~K., \enquote{Adjoint-based shape and kinematics
  optimization of flapping wing propulsive efficiency,} {\it AIAA Paper}
  2013-2472, 2013.
\newblock \doi{10.2514/6.2013-2472},
  \urlprefix\url{https://doi.org/10.2514%2F6.2013-2472}.

\bibitem[{Lee et~al.(2011)Lee, Padulo, and Liou}]{lee2011non}
Lee, B.~J., Padulo, M., and Liou, M.-S., \enquote{Non-sinusoidal trajectory
  optimization of flapping airfoil using unsteady adjoint approach,} {\it AIAA
  Paper} 2011-1312, 2011.
\newblock \doi{10.2514/6.2011-1312},
  \urlprefix\url{https://doi.org/10.2514%2F6.2011-1312}.

\bibitem[{Nadarajah and Jameson(2000)}]{nadarajah2000comparison}
Nadarajah, S., and Jameson, A., \enquote{A comparison of the continuous and
  discrete adjoint approach to automatic aerodynamic optimization,} {\it AIAA
  Paper} 2000-667, 2000.
\newblock \doi{10.2514/6.2000-667},
  \urlprefix\url{https://doi.org/10.2514%2F6.2000-667}.

\bibitem[{Collis et~al.(2001)Collis, Ghayour, Heinkenschloss, Ulbrich, and
  Ulbrich}]{Collis:2001}
Collis, S.~S., Ghayour, K., Heinkenschloss, M., Ulbrich, M., and Ulbrich, S.,
  \enquote{Towards adjoint-based methods for aeroacoustic control,} {\it AIAA
  Paper} 2001-0821, 2001.
\newblock \doi{10.2514/6.2001-821},
  \urlprefix\url{https://doi.org/10.2514%2F6.2001-821}.

\bibitem[{Jameson(2003)}]{jameson2003aerodynamic}
Jameson, A., \enquote{Aerodynamic Shape Optimization Using the Adjoint Method,}
  \emph{VKI Lecture Series on Aerodynamic Drag Prediction and Reduction, von
  Karman Institute of Fluid Dynamics, Rhode St Genese}, 2003, pp. 3--7.

\bibitem[{Moubachir and Zolesio(2006)}]{Moubachir:2006}
Moubachir, M., and Zolesio, J.-P., \emph{Moving shape analysis and control:
  Applications to fluid structure interactions}, Pure and Applied Mathematics,
  Chapman \& Hall/CRC, 2006.
\newblock \doi{10.1201/9781420003246},
  \urlprefix\url{https://doi.org/10.1201%2F9781420003246}.

\bibitem[{Protas and Liao(2008)}]{Protas:2008}
Protas, B., and Liao, W., \enquote{Adjoint-based optimization of PDEs in moving
  domains,} \emph{J.\ Comput.\ Phys.}, Vol. 227, No.~4, 2008, pp. 2707--2723.
\newblock \doi{10.1016/j.jcp.2007.11.014},
  \urlprefix\url{https://doi.org/10.1016%2Fj.jcp.2007.11.014}.

\bibitem[{Xu and Wei(2016)}]{xu2016using}
Xu, M., and Wei, M., \enquote{Using adjoint-based optimization to study
  kinematics and deformation of flapping wings,} \emph{J. Fluid Mech.}, Vol.
  799, 2016, pp. 56--99.
\newblock \doi{10.1017/jfm.2016.351},
  \urlprefix\url{https://doi.org/10.1017%2Fjfm.2016.351}.

\bibitem[{Xu and Wei(2013)}]{XuWei:2013}
Xu, M., and Wei, M., \enquote{Using adjoint-based approach to study flapping
  wings,} {\it AIAA Paper} 2013-839, 2013.
\newblock \doi{10.2514/6.2013-839},
  \urlprefix\url{https://doi.org/10.2514%2F6.2013-839}.

\bibitem[{Xu and Wei(2014)}]{xu2014continuous}
Xu, M., and Wei, M., \enquote{A continuous adjoint-based approach for the
  optimization of wing flapping,} {\it AIAA Paper} 2014-2048, 2014.
\newblock \doi{10.2514/6.2014-2048},
  \urlprefix\url{https://doi.org/10.2514%2F6.2014-2048}.

\bibitem[{Xu et~al.(2015)Xu, Wei, Li, and Dong}]{xu2015adjoint}
Xu, M., Wei, M., Li, C., and Dong, H., \enquote{Adjoint-based optimization of
  flapping plates hinged with a trailing-edge flap,} \emph{Theoretical and
  Applied Mechanics Letters}, Vol.~5, No.~1, 2015, pp. 1--4.
\newblock \doi{10.1016/j.taml.2014.12.005},
  \urlprefix\url{https://doi.org/10.1016%2Fj.taml.2014.12.005}.

\bibitem[{Mittal et~al.(2008)Mittal, Dong, Bozkurttas, Najjar, Vargas, and
  Von~Loebbecke}]{mittal2008versatile}
Mittal, R., Dong, H., Bozkurttas, M., Najjar, F., Vargas, A., and
  Von~Loebbecke, A., \enquote{A versatile sharp interface immersed boundary
  method for incompressible flows with complex boundaries,} \emph{J.\ Comput.\
  Phys.}, Vol. 227, No.~10, 2008, pp. 4825--4852.
\newblock \doi{10.1016/j.jcp.2008.01.028},
  \urlprefix\url{https://doi.org/10.1016%2Fj.jcp.2008.01.028}.

\bibitem[{Li and Dong(2017)}]{li2017wing}
Li, C., and Dong, H., \enquote{Wing kinematics measurement and aerodynamics of
  a dragonfly in turning flight,} \emph{Bioinspiration \& biomimetics},
  Vol.~12, No.~2, 2017, p. 026001.
\newblock \doi{10.1088/1748-3190/aa5761},
  \urlprefix\url{https://doi.org/10.1088%2F1748-3190%2Faa5761}.

\bibitem[{Liu et~al.(2016)Liu, Dong, and Li}]{liu2016vortex}
Liu, G., Dong, H., and Li, C., \enquote{Vortex dynamics and new lift
  enhancement mechanism of wing--body interaction in insect forward flight,}
  \emph{Journal of Fluid Mechanics}, Vol. 795, 2016, pp. 634--651.
\newblock \doi{10.1017/jfm.2016.175},
  \urlprefix\url{https://doi.org/10.1017%2Fjfm.2016.175}.

\bibitem[{Li et~al.(2015)Li, Dong, and Liu}]{li2015effects}
Li, C., Dong, H., and Liu, G., \enquote{Effects of a dynamic trailing-edge flap
  on the aerodynamic performance and flow structures in hovering flight,}
  \emph{Journal of Fluids and Structures}, Vol.~58, 2015, pp. 49--65.
\newblock \doi{10.1016/j.jfluidstructs.2015.08.001},
  \urlprefix\url{https://doi.org/10.1016%2Fj.jfluidstructs.2015.08.001}.

\bibitem[{Xu et~al.(2016)Xu, Wei, Yang, and Lee}]{xu2016embedded}
Xu, M., Wei, M., Yang, T., and Lee, Y.~S., \enquote{An embedded boundary
  approach for the simulation of a flexible flapping wing at different density
  ratio,} \emph{European Journal of Mechanics-B/Fluids}, Vol.~55, 2016, pp.
  146--156.
\newblock \doi{10.1016/j.euromechflu.2015.09.006},
  \urlprefix\url{https://doi.org/10.1016%2Fj.euromechflu.2015.09.006}.

\bibitem[{Xu(2014)}]{xuTheis:2014}
Xu, M., \enquote{Understanding flapping-wing aerodynamics through adjoint-based
  approach,} Ph.D. thesis, New Mexico State University, Las Cruces, NM, 2014.

\bibitem[{Yang(2005)}]{yangTheis:2005}
Yang, J., \enquote{An embedded boundary formulation for Large-Eddy Simulation
  of turbulent flows interacting with moving boundaries,} Ph.D. thesis,
  University of Maryland, College Park, Maryland, 2005.

\bibitem[{Powell(1994)}]{powell1994direct}
Powell, M.~J., \enquote{A direct search optimization method that models the
  objective and constraint functions by linear interpolation,} \emph{Advances
  in optimization and numerical analysis}, Springer, 1994, pp. 51--67.
\newblock \doi{10.1007/978-94-015-8330-5_4},
  \urlprefix\url{https://doi.org/10.1007%2F978-94-015-8330-5_4}.

\bibitem[{Johnson(2014)}]{johnson2014nlopt}
Johnson, S.~G., \enquote{The NLopt nonlinear-optimization package,} , 2014.

\bibitem[{Lauder et~al.(2005)Lauder, Madden, Hunter, Tangorra, Davidson,
  Proctor, Mittal, Dong, and Bozkurttas}]{lauder2005design}
Lauder, G.~V., Madden, P., Hunter, I., Tangorra, J., Davidson, N., Proctor, L.,
  Mittal, R., Dong, H., and Bozkurttas, M., \enquote{Design and performance of
  a fish fin-like propulsor for AUVs,} \emph{Proceedings of 14th International
  Symposium on Unmanned Untethered Submersible Technology, Durham, NH}, 2005.

\bibitem[{Haller(2005)}]{Haller:2005}
Haller, G., \enquote{An objective definition of a vortex,} \emph{J. Fluid
  Mech.}, Vol. 525, 2005, pp. 1--26.
\newblock \doi{10.1017/s0022112004002526},
  \urlprefix\url{https://doi.org/10.1017%2Fs0022112004002526}.

\bibitem[{Wang and Gao(2013)}]{wang2012drag}
Wang, Q., and Gao, J., \enquote{The drag-adjoint field of a circular cylinder
  wake at Reynolds numbers 20, 100 and 500,} \emph{J. Fluid Mech.}, Vol. 730,
  2013, pp. 145--161.
\newblock \doi{10.1017/jfm.2013.323},
  \urlprefix\url{https://doi.org/10.1017%2Fjfm.2013.323}.

\bibitem[{Deng et~al.(2011)Deng, Liu, Zhang, Liu, and Wu}]{deng2011topology}
Deng, Y., Liu, Z., Zhang, P., Liu, Y., and Wu, Y., \enquote{Topology
  optimization of unsteady incompressible Navier--Stokes flows,} \emph{Journal
  of Computational Physics}, Vol. 230, No.~17, 2011, pp. 6688--6708.

\bibitem[{Andersen et~al.(2005)Andersen, Pesavento, and
  Wang}]{andersen2005unsteady}
Andersen, A., Pesavento, U., and Wang, Z., \enquote{Unsteady aerodynamics of
  fluttering and tumbling plates,} \emph{J. Fluid Mech.}, Vol. 541, 2005, pp.
  65--90.
\newblock \doi{10.1017/s002211200500594x},
  \urlprefix\url{https://doi.org/10.1017%2Fs002211200500594x}.

\bibitem[{Floryan et~al.(2017)Floryan, Van~Buren, Rowley, and
  Smits}]{floryan2017scaling}
Floryan, D., Van~Buren, T., Rowley, C.~W., and Smits, A.~J., \enquote{Scaling
  the propulsive performance of heaving and pitching foils,} \emph{Journal of
  Fluid Mechanics}, Vol. 822, 2017, pp. 386--397.
\newblock \doi{10.1017/jfm.2017.302},
  \urlprefix\url{https://doi.org/10.1017%2Fjfm.2017.302}.

\end{thebibliography}

\end{document}